\documentclass[preprint, 11pt, authoryear]{elsarticle}
\usepackage{setspace}
\usepackage[margin=1.2in]{geometry}
\usepackage{url}
\usepackage{pgfplots}
\usepackage{amsmath,amsthm,amssymb}
\usepackage{wasysym}
\pgfplotsset{compat=1.18}
\usepackage{algpseudocodex}
\algrenewcommand\textproc{\texttt}
\usepackage{algorithm}
\usepackage[justification=centering]{caption}
\usepackage[labelformat=simple, justification=centering]{subcaption}
\usepackage{listings}
\usepackage{nameref}
\usepackage{hyperref}
\usepackage[textwidth=28mm]{todonotes}
\usepackage{svg}
\usepackage{soul}
\usepackage{xcolor}
\usepackage{float}
\usepackage{printlen}
\usepackage{tabularx}
\usepackage{mathtools}
\usepackage[bottom]{footmisc}

\setuptodonotes{color=black!10}
\definecolor{highlight}{HTML}{214D7E}

\def\SemiCol{\textbf{; }} 

\biboptions{authoryear}

\journal{European Journal of Operational Research}

\begin{document}
\begin{frontmatter}
	\title{\huge\textbf{An open-source heuristic \\ to reboot 2D nesting research}}

	\author[1]{Jeroen Gardeyn\corref{cor1}}
	\ead{jeroen.gardeyn@gmail.com}

    \author[1]{Greet Vanden Berghe}
	\ead{greet.vanden.berghe@kuleuven.be}
	
	\author[1]{Tony Wauters}
	\ead{tony.wauters@kuleuven.be}
	
	\cortext[cor1]{Corresponding author}
	
	\address[1]{KU Leuven, Department of Computer Science, NUMA, Belgium}
	
	\begin{abstract}
        2D nesting problems rank among the most challenging cutting and packing problems.
        Yet, despite their practical relevance, research over the past decade has seen remarkably little progress.
        One reasonable explanation could be that nesting problems are already solved to near optimality, leaving little room for improvement.
        However, as our paper demonstrates, we are not at the limit after all.
        This paper presents \texttt{sparrow}, an open-source heuristic approach to solving 2D irregular strip packing problems, along with ten new real-world instances for benchmarking.
        Our approach decomposes the optimization problem into a sequence of feasibility problems, where collisions between items are gradually resolved.
        \texttt{sparrow} consistently outperforms the state of the art — in some cases by an unexpectedly wide margin.
        We are therefore convinced that the aforementioned stagnation is better explained by both a high barrier to entry and a widespread lack of reproducibility.
        By releasing \texttt{sparrow}'s source code, we directly address both issues.
        At the same time, we are confident there remains significant room for further algorithmic improvement.
        The ultimate aim of this paper is not only to take a single step forward, but to reboot the research culture in the domain and enable continued, reproducible progress.
    \end{abstract}
	\begin{keyword}
		cutting and packing, irregular, strip packing, open source, nesting
	\end{keyword}
\end{frontmatter}

\section{Introduction}
2D irregular cutting and packing (C\&P) -- or \emph{nesting} -- problems involve fitting a set of smaller items into larger containers, where the shape of the items and/or containers can be irregular (non-rectangular).
Such optimization problems arise in a wide range of industries such as garment manufacturing, furniture production, shipbuilding, printing, laser-cutting, metalworking, woodworking, foam-cutting, leather crafting, and microchip design.
Designing effective algorithms to solve these NP-hard problems \citep{fowler1981optimal} has been an active area of research since the 1960s \citep{art1966approach}.

Given the real-world relevance of 2D irregular C\&P problems and the ongoing shift in ethos toward open-source practices \citep{unesco2021openscience}, one would expect to find a vibrant ecosystem of publicly available implementations that respond to advances in the academic state of the art.
However, despite more than a decade of stagnation -- where there has been little to no improvement in achievable solution quality -- the open-source landscape remains disconnected from academic progress and trails significantly behind in performance.
We hypothesize that there are two interconnected factors primarily responsible for this situation.

The first is a \textbf{high barrier to entry}.
Beyond the fundamental \emph{optimization} challenge of finding good solutions within a vast search space, irregular C\&P problems also pose a significant \emph{geometric} challenge of verifying the feasibility of such solutions in the first place.
Accurately and efficiently determining whether a given placement of items does not involve any collisions is far from trivial.
These compounding challenges are currently holding back academic advances, as all implementations must start from scratch.

The second factor, we believe, that is leading to this stagnation is a \textbf{lack of reproducibility}.
At the time of writing, not a single academically competitive algorithm for 2D irregular C\&P problems has publicly available source code.
Moreover, many methods fail to provide enough explanation to facilitate complete reproducibility.

This paper has two interrelated goals: (i) bridge the gap between academic and open-source algorithms and (ii) lower the barrier to entry for future researchers.
When addressing these goals, we will rely on the recently introduced \emph{Collision Detection Engine} (CDE) for 2D irregular C\&P problems \citep{gardeyn2025decoupling}, which can handle the geometric challenge.
Our entire focus can therefore target the optimization challenge.
On top of the CDE, we will design an open-source heuristic for the 2D irregular strip packing problem, arguably the most canonical 2D nesting problem.
While we focus on strip packing, our broader ambition is to develop an algorithm that can generalize to other variants of 2D irregular C\&P problems without much effort.

{
Throughout this paper, we aim to present our method in an accessible manner, with an emphasis placed on high-level reasoning over exhaustive formal detail.
While implementation details that would hinder readability are intentionally omitted, the accompanying open-source codebase serves as the formal specification of our method, with each algorithm explicitly referenced in the documentation.
This creates two complementary resources: this paper provides intuition and understanding, while the accompanying codebase offers the formal specification for reproduction and implementation.
}

\section{Problem definition} \label{section:spp_definition}

\begin{figure}
    \centering
    \includegraphics[width=0.6\textwidth]{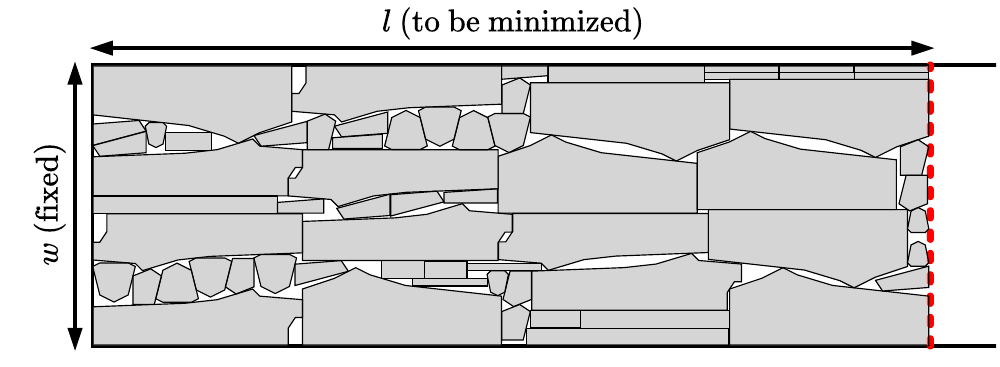}
    \caption{The 2D irregular strip packing problem.}
    \label{fig:strip_packing}
\end{figure}

The 2D irregular strip packing problem (2DISPP) involves placing a set of irregular items into a rectangular strip of fixed width $w$ and variable length $l$.
The objective is to minimize the length of the strip, while ensuring all items are placed entirely within it and none of them collide with each other.
Figure \ref{fig:strip_packing} illustrates the problem through an example solution.

Let us begin by introducing some notation:
\begin{center}
    \begin{tabular}{ll}
        $i \in I$ &: an item $i$ from the set of items to be packed $I$ \\
        $S$ &: the interior of a shape, represented by a set of points $(x,y) \in \mathbb{R}^2$\\
        $S_i$ &: item $i$'s shape \\
        $t$ &: a rigid transformation: a translation $\in \mathbb{R}^2$ +\\
            &\; rotation $\in [0, 2\pi[$ + reflection $\in \{0,1\}$ \\
        $t(S)$ &: the result of transforming shape $S$ by $t$ \\
        $S_a\cap S_b\neq\emptyset$ &: a collision between two shapes \\
    \end{tabular}
\end{center}

We can now formulate the 2DISPP more formally:
\begin{align}
    \text{min} \quad & l \nonumber \\
    \text{s.t.}\quad & t_i(S_i)\subseteq [0,w] \times [0,l] &\forall i\in I \label{eq:collision_strip}\\
    & t_a(S_a)\cap t_b(S_b)=\emptyset &\forall a,b \in I, a\neq b \label{eq:collision_item} \\
    & t_i \in \mathbb{R}^2 \times [0, 2\pi[ \; \times \; \{0,1\} &\forall i\in I \label{eq:transformations} \\
    & l > 0 \label{eq:strip_length}
\end{align}

The objective of the 2DISPP is to determine a set of rigid transformations $\{t_i : i \in I\}$ such that, when each item $i$'s shape is transformed by its corresponding transformation $t_i$, all items are positioned entirely within the strip (Expression \ref{eq:collision_strip}) and none collide with each other (Expression \ref{eq:collision_item}) for the smallest possible strip length $l$.

In practice, the 2DISPP is more restricted than our definition.
Item shapes are usually represented by the interior of simple polygons, with transformations restricted to continuous translations plus a set of discrete rotations (such as multiples of 90 or 180 degrees).
Continuous rotation and reflection (flipping) are seldom considered in academic publications, despite them being relevant for some real-world applications.
In the interest of generality, we aim for our algorithm to be compatible with any rigid transformation.

\section{State of 2D irregular strip packing} \label{section:state_of_the_art}
Solution methods for the 2DISPP fall into three main algorithmic categories. 
We will briefly discuss each category, its notable contributions and some key takeaways.

\subsection*{Exact algorithms}
Exact methods for the 2DISPP typically rely on mathematical programming formulations.
\citet{leao2020irregular} provide a comprehensive survey of the broad range of exact algorithms for irregular C\&P problems.
Such approaches can generally only handle a small set of items, support few angles of rotation and struggle with complex shapes.
Although more recent work -- such as the mixed-integer formulation by \citet{lastra2024mixed} -- demonstrates continued advances, exact approaches remain impractical for many academic instances, let alone complex real-world instances.

\subsection*{Construction-based heuristics}
Construction heuristics build solutions sequentially by placing one item at a time according to a predetermined ordering in addition to a placement rule. 
To improve solution quality, many methods embed a higher-level heuristic that explores different orderings and placement strategies. 
Representative work has been conducted by \citet{oliveira2000topos, gomes2002two, burke2006new, bennell2010beam, amaro2017parallel}.
These algorithms are generally straightforward to implement and reproducible provided that sufficient detail is included in the corresponding publication.
However, the achievable solution quality of construction heuristics lags significantly behind the state of the art.

We can attribute this to the nature of the 2DISPP, which does not lend itself well to such approaches for two main reasons.
First, even minor changes to the ordering of items or their placement early in the construction process leads to unpredictable changes concerning final solution quality.
This chaotic nature undermines the effectiveness of the higher-level heuristics.
Second, if one jumps ahead to the densely packed solutions in Figure \ref{fig:best_solutions}, try to imagine the orderings and placement rules needed to position the items one by one into similar configurations.
Maybe one could devise an effective placement strategy that works well for one of these instances, but it is hard to imagine a single set of rules that would work well for all.

\subsection*{Iterative-improvement heuristics}
Most methods relying on iterative improvement begin with a complete solution and repeatedly apply local modifications -- such as shifting or swapping items -- in an attempt to find new solutions of higher quality.
Early contributions include \citet{egeblad2007fast, umetani2009solving, leung2012extended}. 
More recent advances by \citet{elkeran2013new, wang2017flexible, sato2019raster} represent the state of the art for the 2DISPP.
The common feature shared by these methods is that they tolerate temporary collisions between items.
The strip gets converted into a container with fixed dimensions and, every time a feasible solution is found, its length is reduced further.
This shrinking leads to new collisions, which the aforementioned methods attempt to gradually resolve in order to reach a feasible solution for the shorter strip.

While many of these iterative-improvements heuristics can achieve high-quality solutions, they tend to be more complicated than construction-based heuristics.
Even when their algorithmic flow is presented clearly -- as in \citet{sato2019raster} -- the absence of source code or sufficient implementation details prevents true reproducibility.

\subsection*{Open source}
Solid, robust and publicly available implementations to solve the 2DISPP are scarce.
A notable exception is a tree-search heuristic by \citet{fontan}.
This heuristic has a unique approach to handling the geometric challenge that involves decomposing shapes into trapezoids.  Despite representing a valuable open-source contribution, \citet{fontan}'s method is unrelated to the iterative-improvement methods and cannot consistently compete with the academic state of the art.

\section{A sequence of feasibility problems} \label{section:feasibility_problem}
As indicated in Section \ref{section:state_of_the_art}, the best-performing algorithms for the 2DISPP temporarily relax the non-collision constraint (Expression \ref{eq:collision_item}) during their iterative search.
These methods essentially convert the optimization problem into a \textbf{sequence of feasibility problems}.
Each feasibility problem fixes strip length $l$ and the goal becomes to find a single solution that is feasible. 

\citet{umetani2009solving} referred to this subproblem as the `Overlap Minimization Problem', but we find this term somewhat misleading.
Feasibility, not minimal overlap, is the true goal.
Any solution with overlap is ultimately invalid, regardless of how small it may be.
Additionally, reaching a feasible solution does not entail strictly minimizing overlap, there may be additional factors that aid the search in its goal to eliminate overlap.
Therefore, a more appropriate name for this subproblem would probably be the `Overlap Elimination Problem'.

The idea of exploring infeasible regions has existed for a long time \citep{glover1989tabu} and is a strategy commonly used in the irregular C\&P literature. 
However, the reasons for its effectiveness in nesting problems are rarely elaborated upon. 
We will therefore take this opportunity to offer some intuition for why this conversion works so well.

The 2DISPP is a highly constrained problem, with the non-collision constraint (Expression \ref{eq:collision_item}) being the most significant.
This constraint is so restrictive that virtually any local modification, such as shifting or swapping the positions of items, would immediately render a solution infeasible.
The high-quality solutions we seek are effectively rare oases in a vast desert of infeasibility.
Algorithms that cannot venture outside the feasible region are prevented from making meaningful progress.
The best-performing algorithms therefore temporarily relax this non-collision constraint, enabling them to traverse otherwise inaccessible routes towards high-quality solutions.

\begin{figure}[ht]
    \def\subfigwidth{2.5cm}
    \centering
    \begin{subfigure}[t]{.25\textwidth}
        \centering
        \includegraphics[width=\subfigwidth]{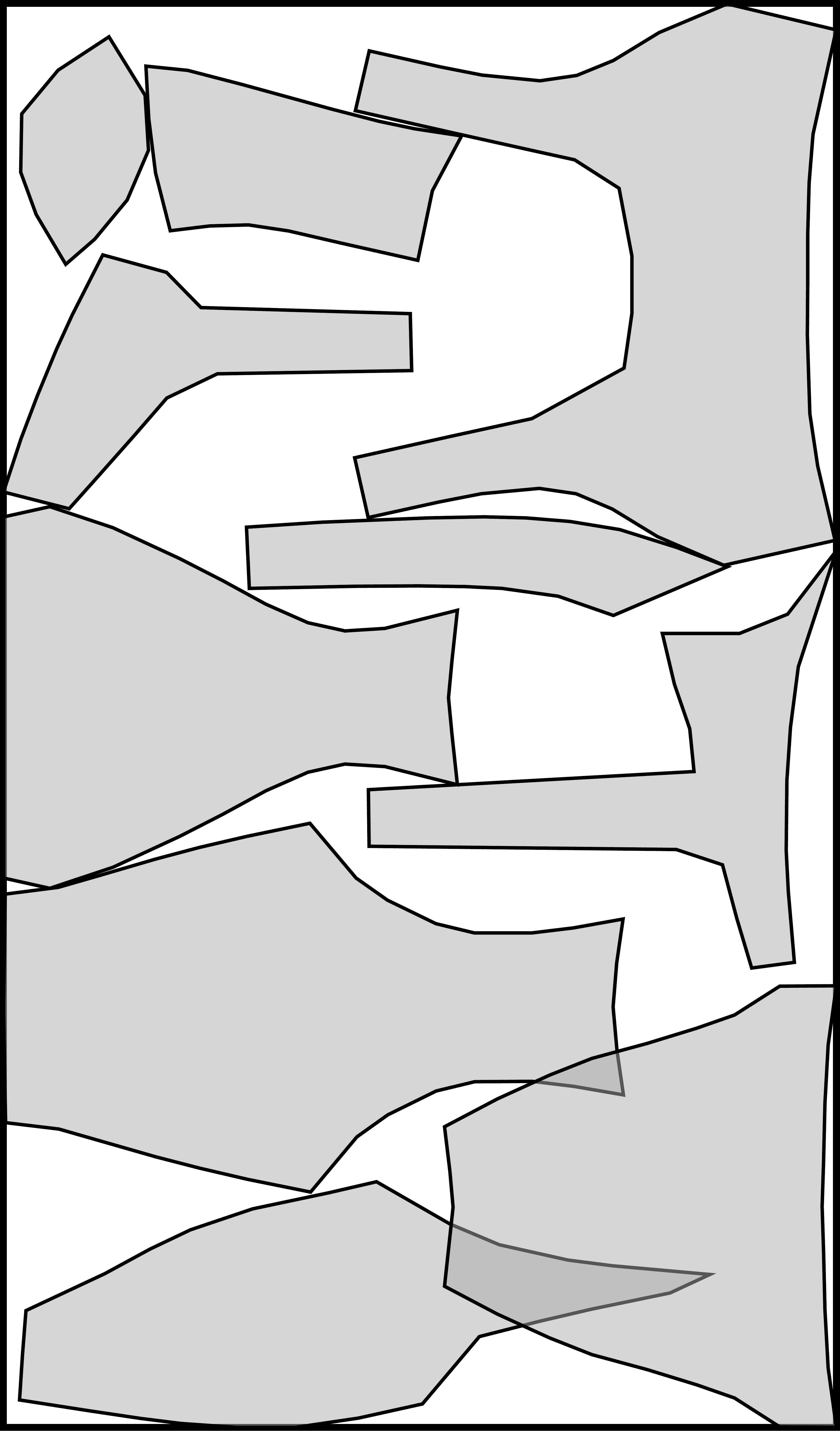}
        \caption{Two colliding pairs of items}
        \label{fig:feasibility_no}
    \end{subfigure}
    \begin{subfigure}[t]{.25\textwidth}
        \centering
        \includegraphics[width=\subfigwidth]{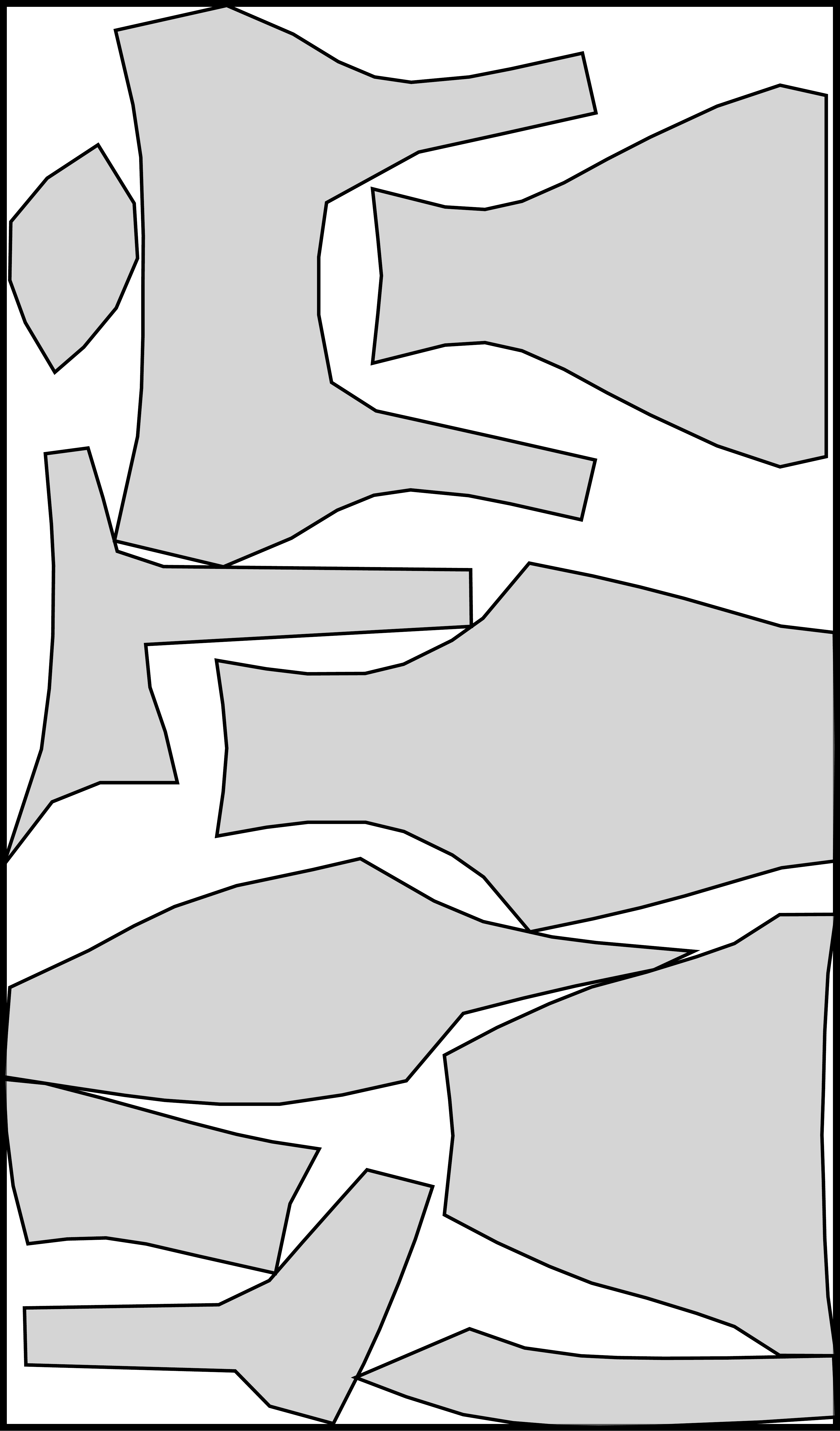}
        \caption{Feasible}
        \label{fig:feasibility_yes}
    \end{subfigure}
    \caption{Example of an infeasible and feasible configuration of a set of items within containers of equal dimensions.}
    \label{fig:feasibility}
\end{figure}
In order to effectively solve the feasibility problem, an algorithm must be able to move from an infeasible solution, such as the one depicted in Figure \ref{fig:feasibility_no}, towards a feasible solution like the one in Figure \ref{fig:feasibility_yes}.
We identify three key components required to achieve this:

\begin{itemize}
    \item \textbf{Assessing feasibility}: 
    A binary collision check to determine which constraints are violated.
    \item \textbf{Search algorithm}: 
    A strategy that explores the solution space in order to incrementally resolve collisions.
    \item \textbf{Quantifying infeasibility}: 
    As collisions are often not resolvable in a single step, a binary check is too coarse to effectively guide the search. A smooth, continuous metric is required to evaluate the severity of each collision and direct the search algorithm.
\end{itemize}

Sections \ref{section:detecting_collisions}-\ref{section:solve_feas} introduce these components one by one and eventually combine them into a complete algorithm to solve the feasibility problem.
Finally, Section \ref{section:spp_heuristic} will integrate everything into a heuristic to solve the 2DISPP.

\section{Detecting collisions} \label{section:detecting_collisions}
The most elementary challenge involved in solving any feasibility problem is assessing whether or not a constraint is violated.
Given the geometric challenge that accompanies nesting problems, checking the feasibility of a solution or even of a single placement in a way that is efficient, precise and robust is not trivial.
The fundamental ways of dealing with this geometric challenge have been described by \citet{bennell2008geometry}.
Currently, all state-of-the-art nesting algorithms rely on either \emph{no-fit polygons} (NFPs) or a \emph{raster} to handle the geometric challenge.
However, these two approaches have serious limitations regarding robustness and precision, respectively.

A third approach mentioned by \citet{bennell2008geometry} is trigonometry: a simple, precise, robust and general way of detecting collisions between two polygons.
This approach was historically very sensitive to the total number of edges in the shapes involved, and therefore seldom used in practice.

\citet{gardeyn2025decoupling} introduced a \emph{Collision Detection Engine} (CDE) to provide a fast and reliable way for 2D nesting algorithms to assess feasibility.
This CDE builds on trigonometric principles and includes a comprehensive suite of algorithmic enhancements to overcome the traditional limitations of such trigonometric approaches.
The CDE is implemented in an open-source library called \texttt{jagua-rs}\footnote{\url{https://github.com/JeroenGar/jagua-rs}}.
\texttt{jagua-rs} supports irregular shapes for both items and containers, continuous rotation and translation, and easily extends to problem variants containing additional spatial constraints.

The optimization algorithm we will introduce in this paper, is built on top of \texttt{jagua-rs}: assessing feasibility through collision \textbf{queries} and issuing \textbf{updates} to reflect any changes in the solution.
All interactions with \texttt{jagua-rs} throughout this paper can be abstracted into the following two functions:
\begin{flushleft}
    \begin{tabular}{@{}l@{}} 
        \texttt{jagua-rs::collisions}$(S) \rightarrow C$ \\
        \quad Returns the set of items $C$ $(\subseteq I)$ that collide with shape $S$. 
        \\[1ex]
        \texttt{jagua-rs::move\_item}$(i, t)$ \\
        \quad Moves item $i$ to a new position defined by transformation $t$.
    \end{tabular}
\end{flushleft}

\section{Quantifying collisions} \label{section:quantifying_collisions}

To facilitate an effective search algorithm, the binary feasibility check provided by the CDE (Section \ref{section:detecting_collisions}) needs to be converted into a continuous one capturing the \emph{severity of the collision}: an expression of the expected difficulty to resolve it.

\begin{algorithm}[ht]
    \caption{\texttt{evaluate\_item\_pair}$(a,b)$}
    \begin{algorithmic}[1]
        \If{$a \in \Call{jagua-rs::collisions}{S_b}$}
        \State \Return $\Call{quantify\_collision}{S_a,S_b}$
        \Comment{Alg. \ref{alg:quantify_collision}}
        \Else
        \State \Return 0
        \EndIf
    \end{algorithmic}
    \label{alg:eval_pair}
\end{algorithm}

Let us begin by introducing Algorithm \ref{alg:eval_pair}, the general function to evaluate any pair of items $a,b \in I$. 
The procedure begins by querying the CDE to determine whether the two items are colliding with each other.
When no collision occurs, the function returns 0.
When a collision is detected, the shapes of the items involved will be passed to the \nameref{alg:quantify_collision} function.

Many existing approaches for the 2DISPP quantify the severity of a collision in some way.
\citet{imamichi2009iterated} proposed using \emph{penetration depth}, which denotes the smallest translation needed to separate two colliding entities.
Instead of the exact penetration depth, they measure the minimum translation required along a single axis, such as horizontally or vertically. 
\citet{leung2012extended}, \citet{elkeran2013new}, \citet{wang2017flexible} and \citet{sato2019raster} each employ some variant of penetration depth to quantify the severity of a collision.
Regardless of the concrete implementation, these penetration-depth derivatives are essentially all one-dimensional metrics.

When relying on a trigonometric approach to detect collisions, such as \texttt{jagua-rs}, computing penetration depth is impractical.
We therefore need to find a different way to efficiently quantify the severity of a collision. 
It is important to stress that this function will only serve to guide the search towards a feasible solution as effectively as possible.
It does not need to represent anything tangible, nor does it need to be precise.
We aim for a metric that supports continuous rotation and is insensitive to the complexity of the shapes involved, for the sake of greater generality and real-world applicability.

The remainder of this section describes how we design our quantification function, eventually leading to the complete \nameref{alg:quantify_collision} function that is called within Algorithm \ref{alg:eval_pair}.

\subsection{Area of overlap}
\begin{figure}[h]
    \centering
    \begin{subfigure}{.3\textwidth}
        \centering
        \includegraphics[height=0.4\textwidth]{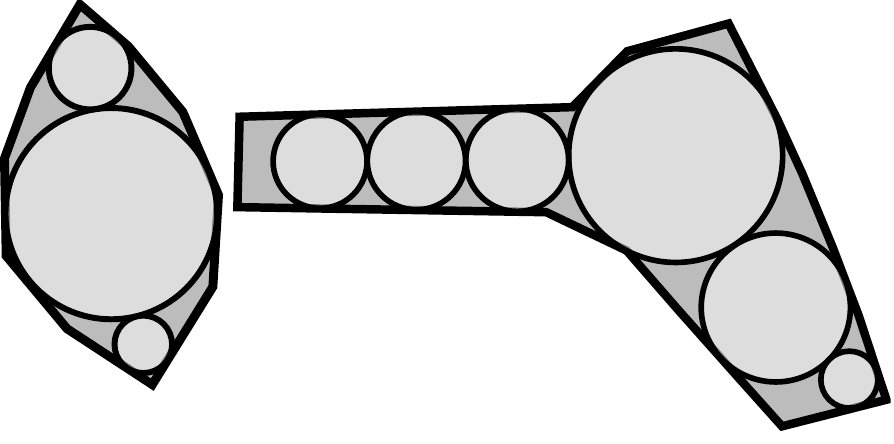}
        \caption{Separated}
        \label{fig:poles_no_overlap}
    \end{subfigure}
    \begin{subfigure}{.3\textwidth}
        \centering
        \includegraphics[height=0.4\textwidth]{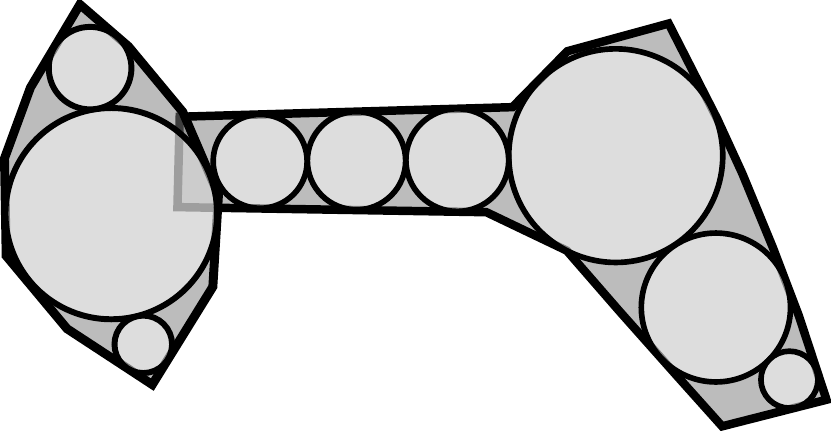}
        \caption{Collision (minor)}
        \label{fig:poles_minor_overlap}
    \end{subfigure}
    \begin{subfigure}{.3\textwidth}
        \centering
        \includegraphics[height=0.4\textwidth]{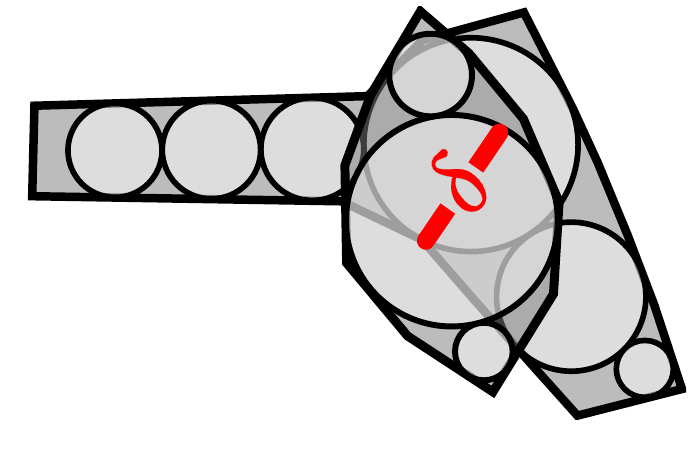}
        \caption{Collision (major)}
        \label{fig:poles_overlap}
    \end{subfigure}
    \caption{Two shapes in different configurations, with their respective poles drawn.
    In (c), the penetration depth ($\delta$) between the largest pair of poles is marked in red.}
    \label{fig:poles}
\end{figure}

Figure \ref{fig:poles} shows two shapes in three different placements, once separated and twice colliding.
Intuitively, the collision in Figure \ref{fig:poles_overlap} appears harder to resolve than the one in Figure \ref{fig:poles_minor_overlap}.
The intersection area -- how much the shapes overlap -- is an obvious candidate for measuring collision severity.
However, calculating this metric precisely is far too expensive to be practical in an iterative improvement heuristic.
Let us investigate whether we can use a shape's \emph{pole of inaccessibility} to devise a cheap proxy.

A shape's pole of inaccessibility is the point within its interior that lies furthest from the boundary.
This point, along with its distance to the boundary, defines the largest inscribable circle in the shape.
Figure \ref{fig:poles} shows a sequence of such inscribed circles -- referred to simply as \emph{poles} from now on -- for each shape. \citet{gardeyn2025decoupling} describe the precise procedure to compute these sets of poles, extending the approach by \citet{Agafonkin_Polylabel_a_fast_2016}.
The CDE leverages them to accelerate the collision detection process and lower its sensitivity to the complexity of the shapes.
The set of poles of shape $S$ is represented by $P(S)$.
Poles are generated once (in preprocessing) for every base shape and can be efficiently transformed in unison with any rigid transformation applied to the shape.

\begin{algorithm}
    \caption{\texttt{overlap\_proxy}($S_a,S_b$)}
    \begin{algorithmic}[1]
        \For{$(p_a, p_b) \in P(S_a) \times P(S_b)$} \Comment{all pairs of poles}
            \State $\delta \gets \textsl{penetration depth}(p_a,p_b)$
            \If{$\delta > 0$}
                \State $\alpha \gets \alpha + \delta \cdot \text{min}\{\diameter(p_a),\diameter(p_b)\}$ \Comment{$\diameter$: diameter}
            \EndIf
        \EndFor
        \State \Return $\alpha$
    \end{algorithmic}
    \label{alg:overlap_area_proxy}
\end{algorithm}

Algorithm \ref{alg:overlap_area_proxy} introduces a proxy for the overlap between two shapes that leverages these poles.
For every combination of poles from the two shapes, the penetration depth ($\delta$) is calculated as the sum of their radii minus the distance between their centers.
All these penetration depths are then weighted by the diameter of the smaller pole and combined into a weighted sum $\alpha$, which provides a suitable proxy for the area of overlap.

{
  \newcommand{\colorgradientbox}{%
    \tikz[baseline=-0.5ex]{
      \node[minimum height=1em, minimum width=3em, draw=black, line width=0.8pt,
            shading=horizontal, left color=white, right color=red] (char) {};
    }\;%
  }

  \newcommand{\solidcolorbox}{%
    \tikz[baseline=-0.5ex]{
      \node[minimum height=1.0em, minimum width=1.0em, draw=black, line width=0.8pt, fill=green!20!white] (char) {};
    }\;%
  }

  \begin{figure}[h]
      \centering
      \begin{subfigure}{.33\textwidth}
          \centering
          \includegraphics[trim={60 100 220 180}, clip=true, width=1.0\textwidth]{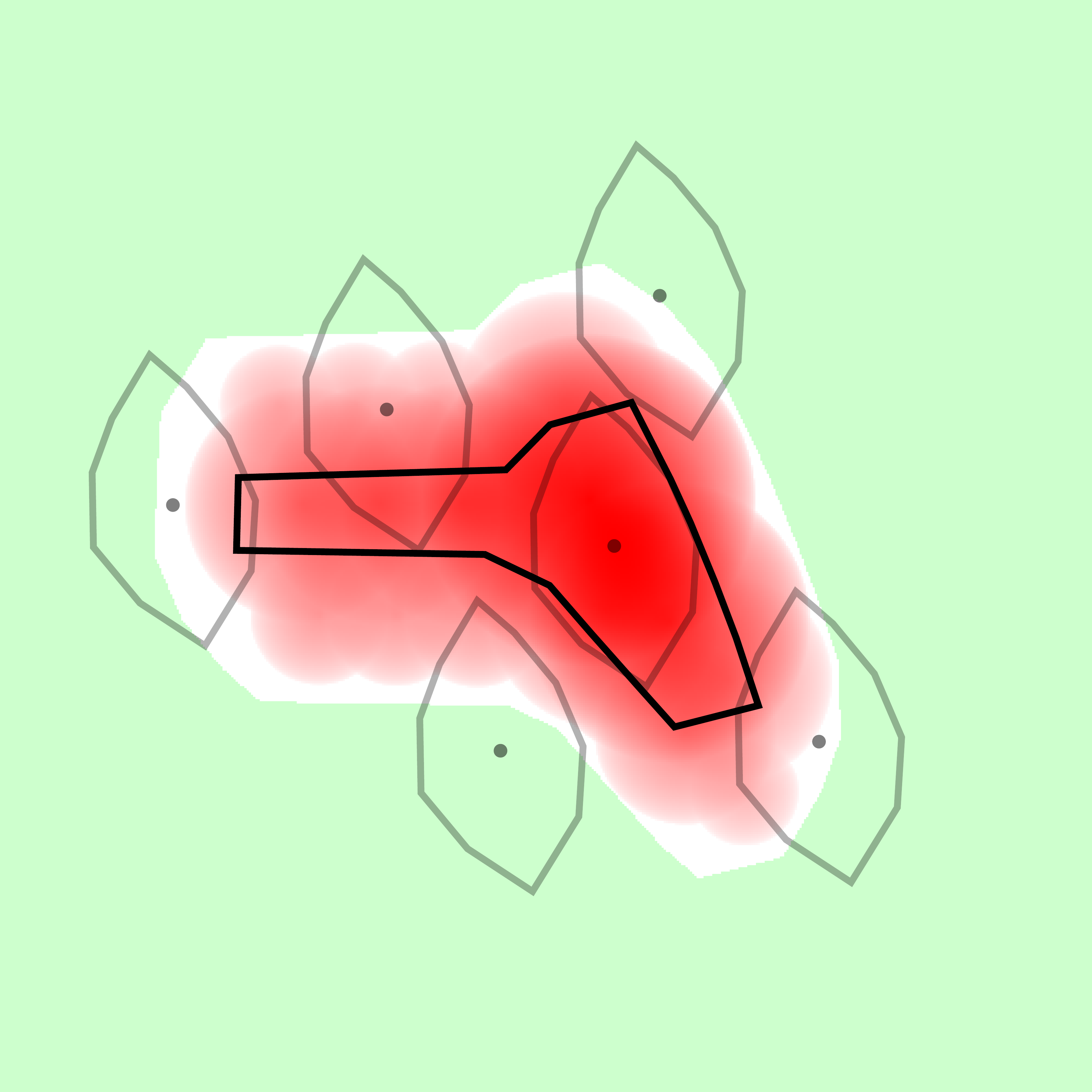}
          \caption{2D}
          \label{fig:overlap_no_decay_2d}
      \end{subfigure}
      \hspace{0.1\textwidth}
      \begin{subfigure}{.33\textwidth}
          \centering
          \includegraphics[trim={600 400 600 800}, clip=true, width=1.0\textwidth]{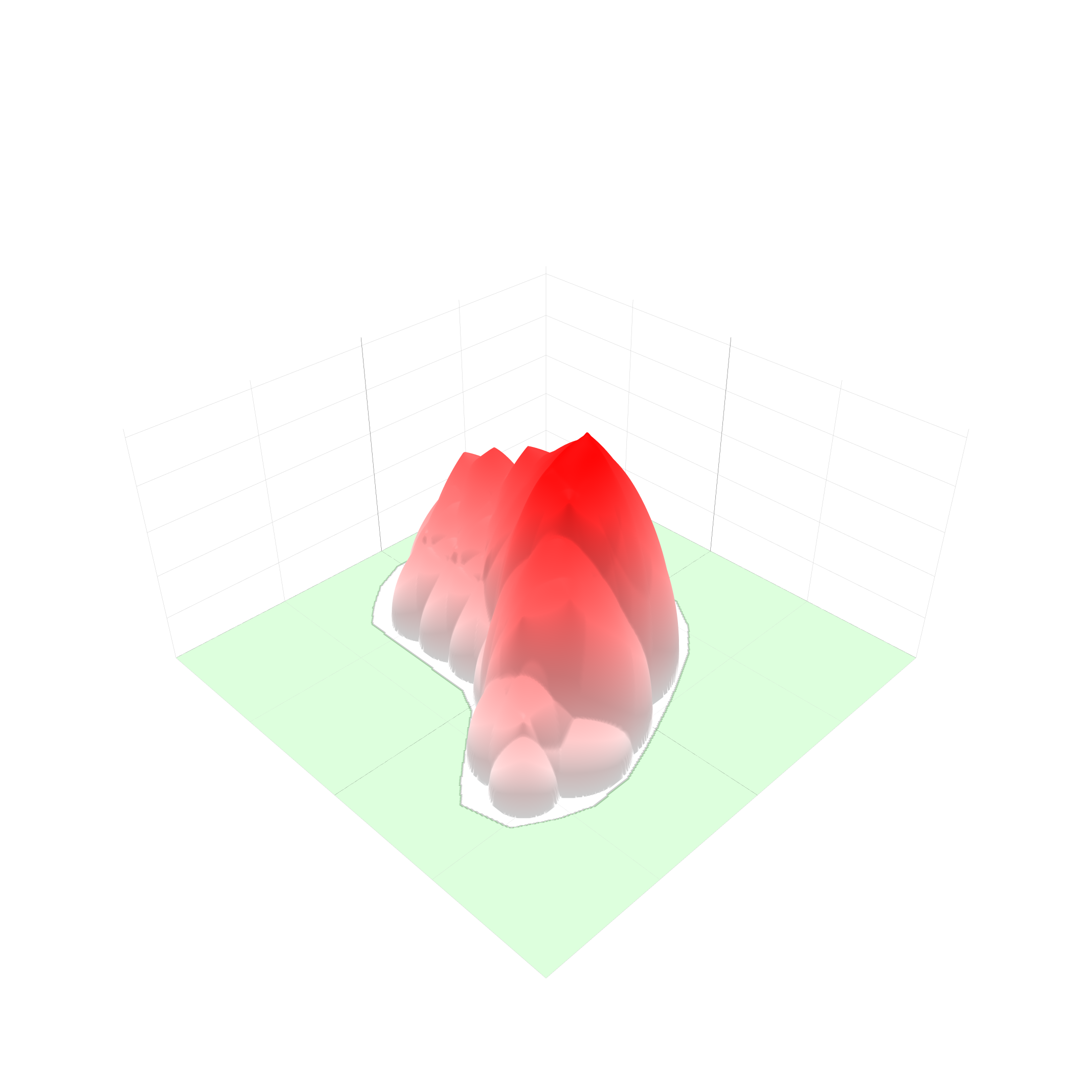}
          \caption{3D}
          \label{fig:overlap_no_decay_3d}
      \end{subfigure}
      \caption{Visualization of the overlap proxy for the same pair of items from Figure~\ref{fig:poles}.
      {\protect\solidcolorbox}: feasible, {\protect\colorgradientbox}: \nameref{alg:overlap_area_proxy}.}
      \label{fig:overlap_no_decay}
  \end{figure}
}

Figure \ref{fig:overlap_no_decay_2d} visualizes the output of this function for the pair of items from Figure \ref{fig:poles}, with the smaller item being positioned all around the larger one. 
The shapes and reference points of six example placements are shown.
The region where the two items do not collide is colored green.
The region where they collide is shaded red, with the intensity of the red mapping to the output of the function described in Algorithm \ref{alg:overlap_area_proxy}.
To help interpret the intensity further, Figure \ref{fig:overlap_no_decay_3d} maps the value of this function to the height of a 3D surface.

In theory, this proxy function has a quadratic complexity $O(n^2)$ regarding the number of poles.
In practice however, the computation time is manageable because (i) a relatively small number of poles (8-16) suffices to capture most shapes and (ii) the loop can be vectorized in modern CPUs either automatically by the compiler or by explicitly leveraging SIMD\footnote{Single Instruction, Multiple Data} instructions.

This overlap proxy exhibits many desirable properties of a quantification function for the severity of a collision:
fast computation, a smooth gradient, and a value that consistently increases with the amount of overlap.
However, poles do not cover the entire item and therefore collisions may occur without any pairs of poles overlapping.
For the collision depicted in Figure \ref{fig:poles_minor_overlap} or throughout the entire white region in Figure \ref{fig:overlap_no_decay}, Algorithm \ref{alg:overlap_area_proxy} will return 0 even though the CDE rightfully detects a collision.
This is problematic because these (minor) collisions are not contributing anything to the quantification function and consequently provide no information to help guide the search.

\subsection{Decaying penetration depth}
To address the aforementioned issue, we introduce a \emph{decaying} version of the penetration depth between pairs of poles.
When $\delta$ is smaller than a certain threshold $\varepsilon$, we switch to a hyperbolic decay function that asymptotically approaches 0 as the distance between the pair of poles increases.
The decaying penetration depth $\delta'$ is defined as follows:
\begin{equation}
    \delta' = \begin{cases}
        \delta & \text{if } \delta > \varepsilon \\
        \frac{\varepsilon^2}{-\delta + 2\varepsilon} & \text{otherwise}
    \end{cases}
    \label{eq:pd_decay}
\end{equation}
A comparison of the decaying variant and standard penetration depth function profiles is visualized in Figure \ref{plot:penetration_base_decay}.

\usepgfplotslibrary{fillbetween}
\pgfplotsset{compat=1.18}

\newcommand{\eps}{1}

\definecolor{purple}{rgb}{255,0,255}
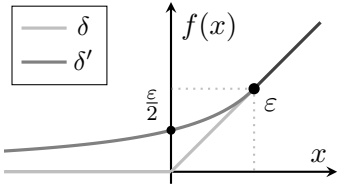
\begin{figure}[h]
    \centering
    \begin{tikzpicture}
        \begin{axis}[
            xlabel={$x$},
            ylabel={$f(x)$},
            xmin=-2,
            xmax=2,
            ymin=-0.1,
            ymax=2,
            axis lines=middle,
            width=6cm,
            height=4cm,
            axis equal,
            axis line style={thick},
            xtick=\empty,
            ytick=\empty,
            grid=both,
            legend style={
                at={(0.02,0.98)},  
                anchor=north west,
                draw=black,         
                fill=none,         
                font=\small        
            },
        ]

        \addplot[
            domain=0:\eps,
            samples=10,
            color=black!25,
            very thick,
            opacity=1.0,
        ] {x};
        \addlegendentry{$\delta$}

        \addplot[
            domain=-5:\eps,
            samples=100,
            color=black!50,
            very thick,
        ]{(\eps)^2 / (-x + 2 * \eps)};
        \addlegendentry{$\delta'$}

        \addplot[
            domain=-5:0,
            samples=10,
            color=black!25,
            very thick,
            opacity=1.0
        ] {0};

        \addplot[
            domain=\eps:1.8,
            samples=10,
            color=black!75,   
            very thick,          
        ] {x};     


        \addplot[
            thick, 
            samples=10, 
            domain=0:6,
            gray, 
            opacity=0.5,
            dotted,
            name path=three
        ] coordinates {(0, \eps)(\eps,\eps)(\eps,0)};
        
        \node[circle, fill=black, inner sep=1.5pt] at (axis cs:\eps,\eps) {};
        \node[below right] at (axis cs:\eps,\eps) {$\varepsilon$};

        \node[
            circle, 
            fill=black, 
            inner sep=1.2pt
            ] at (axis cs:0,\eps / 2) {};
        \node[
            above left, 
            black 
        ] at (axis cs:0,\eps / 2) {$\frac{\varepsilon}{2}$};
    \end{axis}
    \end{tikzpicture}
    \caption{Comparing the standard ($\delta$) and decaying ($\delta'$) penetration depth functions.}
    \label{plot:penetration_base_decay}
\end{figure}

\begin{algorithm}[h]
    \caption{\texttt{overlap\_proxy\_decay}($S_a,S_b$)}
    \begin{algorithmic}[1]
        \State $\varepsilon \gets \texttt{R}_{\varepsilon} \cdot \text{max}\{\diameter({S_a}), \diameter({S_b})\}$ \Comment{$\diameter$: diameter}
        \For{$(p_a, p_b) \in P(S_a) \times P(S_b)$}
            \State $\delta \gets \textsl{penetration depth}(p_a,p_b)$
            \If{$\delta > \varepsilon$}
                 \State $\delta' \gets \delta$
            \Else
                \State $\delta' \gets \varepsilon^2 / (-\delta + 2\varepsilon)$
            \EndIf
            \State $\alpha \gets \alpha + \delta' \cdot \text{min}\{\diameter(p_a),\diameter(p_b)\}$
        \EndFor
        \State \Return $\alpha$
    \end{algorithmic}
    \label{alg:overlap_area_proxy_decay}
\end{algorithm}

{
  \newcommand{\colorgradientbox}{%
    \tikz[baseline=-0.5ex]{
      \node[minimum height=1em, minimum width=3em, draw=black, line width=0.8pt,
            shading=horizontal, left color=white, right color=red] (char) {};
    }\;%
  }

  \newcommand{\solidcolorbox}{%
    \tikz[baseline=-0.5ex]{
      \node[minimum height=1.0em, minimum width=1.0em, draw=black, line width=0.8pt, fill=green!20!white] (char) {};
    }\;%
  }

  \begin{figure}[htbp]
    \centering    
    \begin{subfigure}{.33\textwidth}
        \centering
        \includegraphics[trim={60 100 220 180}, clip=true, width=1.0\textwidth]{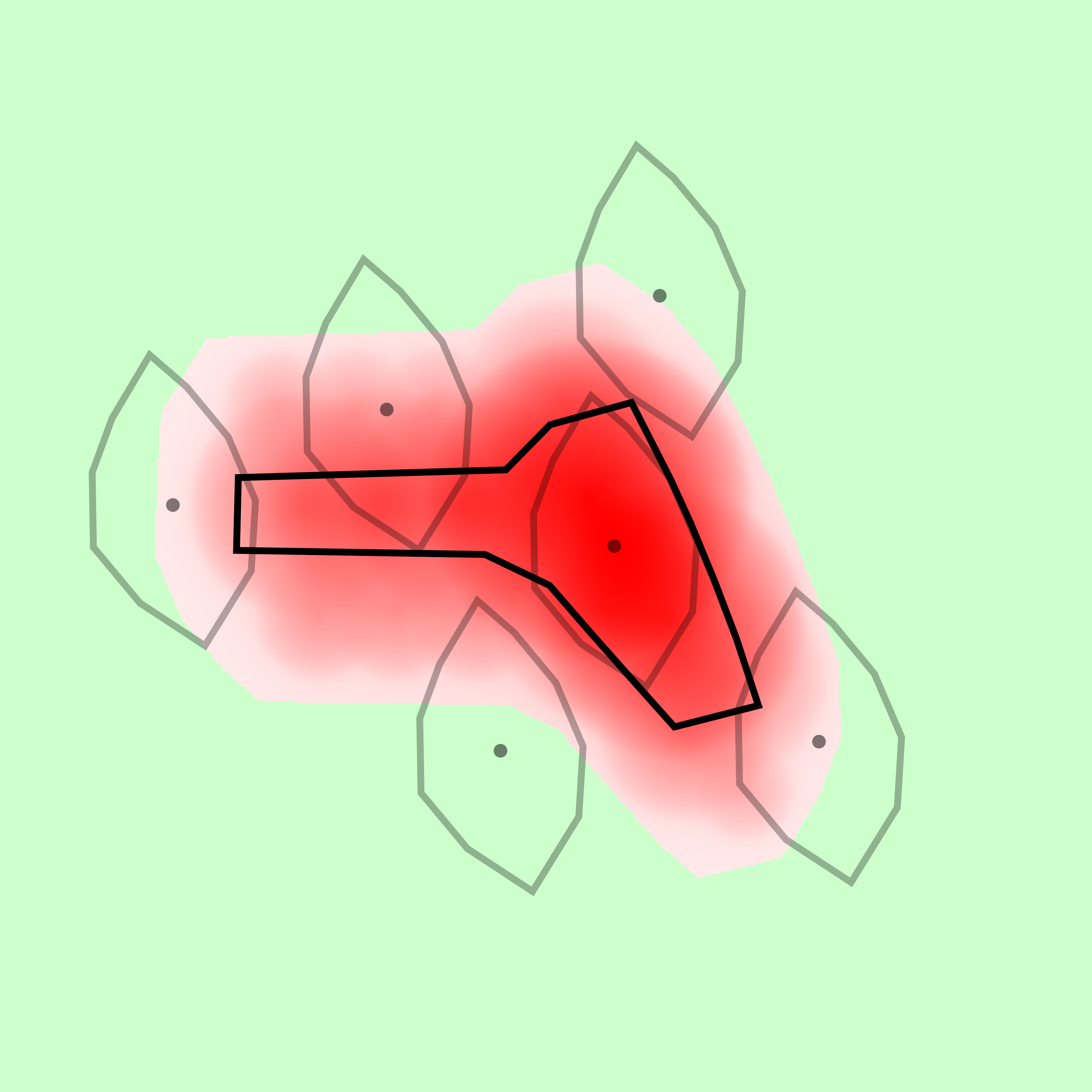}
        \caption{2D}
    \end{subfigure}
    \hspace{0.1\textwidth}
    \begin{subfigure}{.33\textwidth}
        \centering
        \includegraphics[trim={600 400 600 800}, clip=true, width=1.0\textwidth]{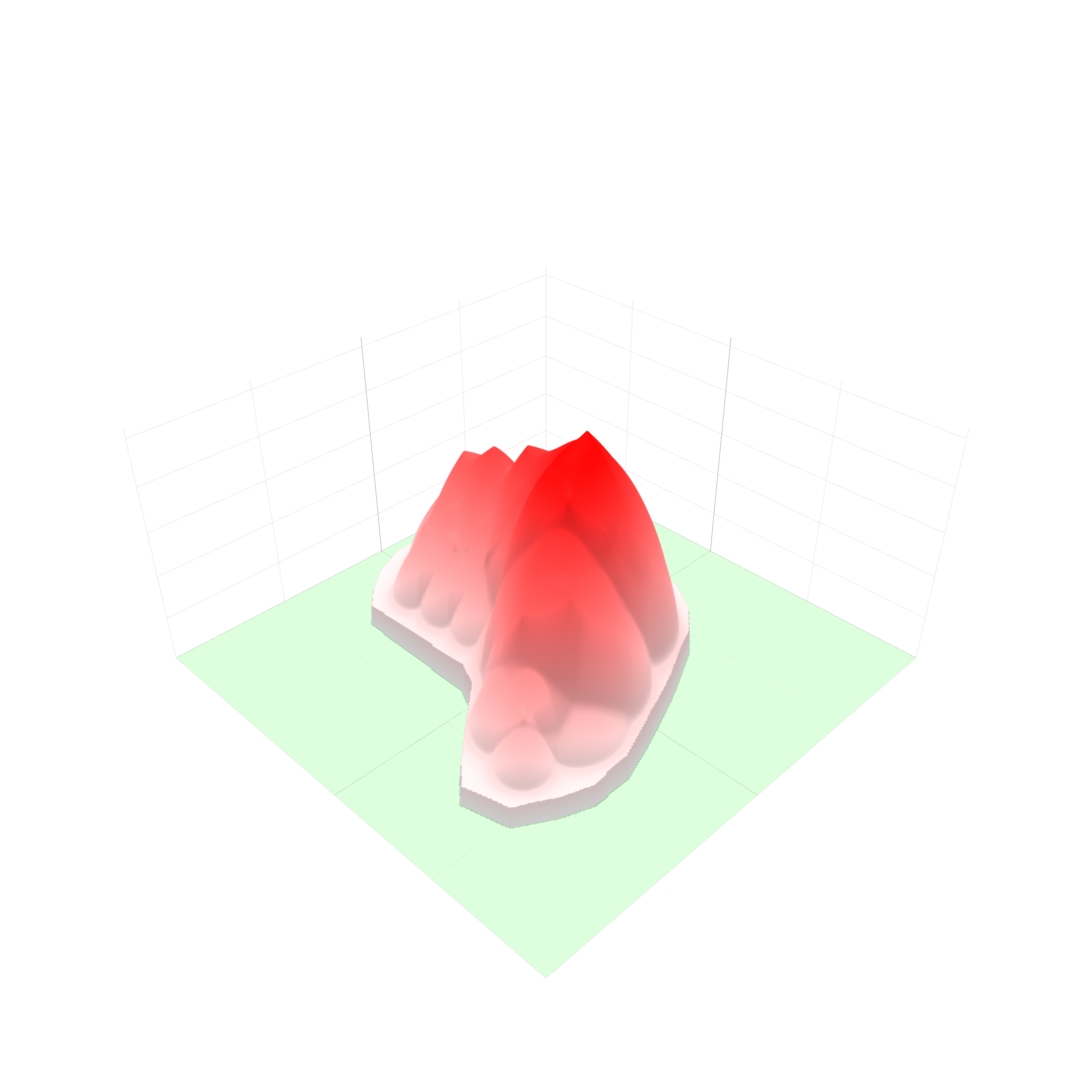}
        \caption{3D}
    \end{subfigure}
    \caption{Visualization of the decaying overlap proxy for the same pair of items from Figure~\ref{fig:poles}.
    {\protect\solidcolorbox}: feasible, {\protect\colorgradientbox}: \nameref{alg:overlap_area_proxy_decay}.}
    \label{fig:overlap_decay}
  \end{figure}
}

Algorithm \ref{alg:overlap_area_proxy_decay} describes the modified overlap proxy, which employs the decaying penetration depth.
Threshold $\varepsilon$ is computed as the larger diameter of the shapes -- the furthest distance between any two of its points -- multiplied by a small constant $\texttt{R}_\varepsilon$.
Figure \ref{fig:overlap_decay} visualizes the new function in the same way as Figure \ref{fig:overlap_no_decay}.

Every collision now contributes to the quantification function and there is a smooth gradient present everywhere.
Additionally, a sharp drop-off is now present at the feasibility boundary, resulting in a significant distinction between a colliding and a non-colliding placement.

\subsection{Shape-based penalty}
The proxy provided by Algorithm \ref{alg:overlap_area_proxy_decay} is a decent starting point for quantifying the severity of a collision.
However, the area of overlap is not necessarily the only factor that influences how difficult a collision is to resolve.
When quantifying collisions purely on the basis of the amount of overlap, we noticed that our search algorithm had a strong tendency to converge to situations with minor amounts of overlap between pairs of items with large and/or concave shapes.
It seems that resolving such collisions is disproportionately more disruptive to the structure of the overall solution and therefore much harder compared to resolving collisions between smaller and/or more convex items.

To counter this, we experimented with adding a fixed term $\lambda$, that penalizes collisions involving such troublesome shapes.
One candidate to base the penalty on is the area of a shape's convex hull, given that it captures both the size and the concavity of a shape.
However, employing this area directly penalized these troublesome shapes too excessively. 
The square root of this area instead provides a much better balance. Therefore, we define shape $S_a$'s penalty $\lambda_{a}$ as:
\begin{equation} \label{eq:lambda}
    \lambda_{a} = \sqrt{\textsl{area}(\textsl{convex hull}(S_a))}
\end{equation}

Since a collision always takes place between a pair of shapes, we need to merge individual penalties.
From our experiments, we found that the geometric mean of the two penalties was the most effective way to combine them.
For a collision between $S_a$ and $S_b$, we define the combined penalty as:
\begin{equation} \label{eq:lambda_ab}
    \lambda_{ab} = \sqrt{\lambda_{a} \cdot \lambda_{b}}
\end{equation}

\begin{algorithm}
    \caption{\texttt{quantify\_collision}($S_a,S_b$)}
    \begin{algorithmic}[1]
        \State $\alpha \gets \text{\nameref{alg:overlap_area_proxy_decay}}$ \Comment{Alg. \ref{alg:overlap_area_proxy_decay}}
        \State $\lambda_{ab} \gets \sqrt{\lambda_a \cdot \lambda_b}$    \Comment{Eq. \ref{eq:lambda_ab}}
        \State \Return $\sqrt{\alpha} \cdot \lambda_{ab}$
    \end{algorithmic}
    \label{alg:quantify_collision}
\end{algorithm}

Algorithm \ref{alg:quantify_collision} presents the full procedure for quantifying a collision between two shapes.
We first compute overlap proxy $\alpha$, and then scale its square root by the combined penalty $\lambda_{ab}$.
Taking the square root of $\alpha$ converts the proxy from an area-like (2D) metric into a one-dimensional one, making it more comparable to the penetration depth derivatives commonly used in other approaches.
This function can now be used in Algorithm \ref{alg:eval_pair} to complete the evaluation of pairs of items.

\section{Navigating a continuous search space} \label{section:continuous_search_space}

All that now remains is to design a search algorithm that is able to gradually resolve collisions.
One of the most fundamental components of any local search algorithm is the \emph{move} operation, enabling the algorithm to transition from one solution to a neighboring one.
Algorithm \ref{alg:move_items} introduces the local search move our approach will employ.
Its core idea -- repositioning every colliding item, one by one, in a randomized order -- is shared by \citet{umetani2009solving,sato2019raster,elkeran2013new}.

\begin{algorithm}[ht]
    \caption{\texttt{move\_items}}
    \begin{algorithmic}[1]
        \State $I_c \gets \{i \in I : |\Call{jagua-rs::collisions}{S_i}| > 1 \}$
        \Comment{colliding items}
        \For{$i \in {I_c}$ in \textbf{a random order}}
            \State $t \gets$ \nameref{alg:search_position} \Comment{Alg. \ref{alg:search_position}}
            \State \Call{jagua-rs::move\_item}{$i,t$}
        \EndFor
    \end{algorithmic}
    \label{alg:move_items}
\end{algorithm}

In 2D irregular C\&P problems, items can be positioned in a continuous 2D space and sometimes also rotated in a continuous manner.
This means it is far too time-consuming to evaluate a full neighborhood of all possible item positions.
Previous approaches have handled this continuous search space in a variety of ways.
\citet{sato2019raster} discretize the search space into a (multi-resolution) grid.
Meanwhile, approaches that rely on NFPs \citep{umetani2009solving,elkeran2013new} narrow the search space down to a set of positions along the edges of the union of all NFPs.
From this (still continuous) set of positions, only a few specific points such as the vertices of the combined NFP are typically considered.
All of these approaches effectively \emph{restrict} the search space to a finite set of positions.

Instead of restricting the search space, we will preserve its continuous nature by relying on a \emph{sampling} strategy to search for new item positions.
Such an approach only evaluates a limited number of sampled positions, but every position remains a possible candidate.
The aim is to obtain high-quality placements while keeping the number of samples to be evaluated as low as possible.

\begin{algorithm}[ht]
    \caption{\texttt{search\_position}($i$)}
    \begin{algorithmic}[1]
        \State $f(i,t)\mapsto e \ \gets$ \nameref{alg:eval_sample} \Comment{Alg. \ref{alg:eval_sample}}
        \State $t^* \gets \Call{perform\_sampling}{i, f(i,t) \mapsto e}$ \Comment{Fig. \ref{fig:sampling}}
        \State \Return $t^*$
    \end{algorithmic}
    \label{alg:search_position}
\end{algorithm}

Algorithm \ref{alg:search_position} indicates how we search for a new position (transformation) for an item $i$.
A sample is defined as an item and an accompanying transformation: $(i, t)$.
We define a function to evaluate samples $f(i,t) \mapsto e$ which will, for now, remain abstract: it takes as input a sample and returns a single value $e$ indicating the quality of that sample.
This function, along with the item, is passed to the sampling procedure, which returns $t^*$, the best position it found to reposition item $i$ to.

Thanks to the open-source nature of our implementation, we can visually illustrate how the \texttt{perform\_sampling} function inside Algorithm \ref{alg:search_position} works.
This allows us to provide an intuitive understanding of the process without relying on convoluted pseudocode or dense mathematical notation.
Readers interested in the concrete implementation details can refer directly to the source code (Section \ref{section:implementation}).

\begin{figure}[htbp]
    \def\subfigheight{4.5cm}
    \centering
    \begin{subfigure}[t]{.19\textwidth}
        \centering
        \includegraphics[height=\subfigheight]{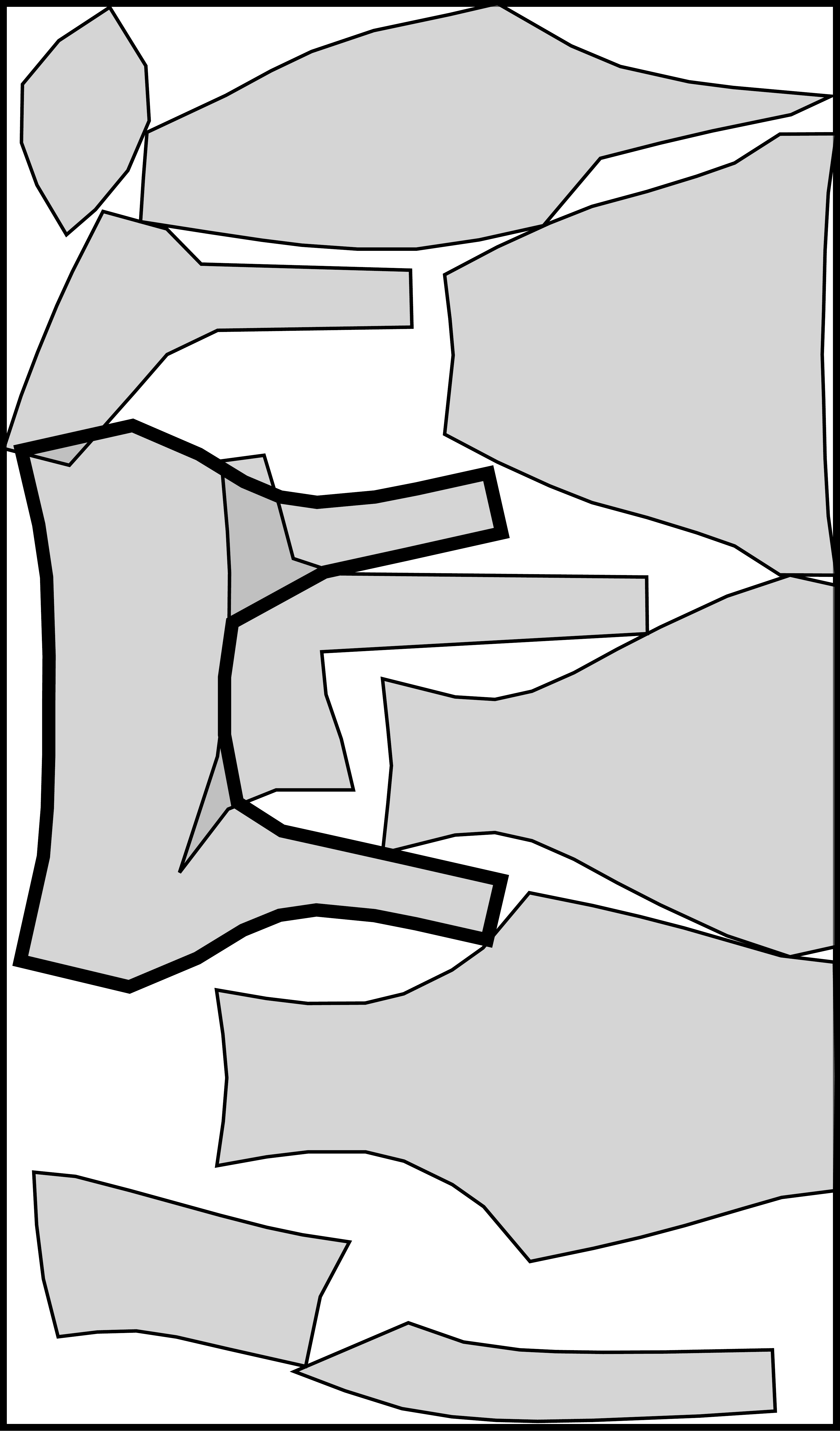}
        \caption{Before}
        \label{fig:sampling_before}
    \end{subfigure}
    \begin{subfigure}[t]{.19\textwidth}
        \centering
        \includegraphics[height=\subfigheight]{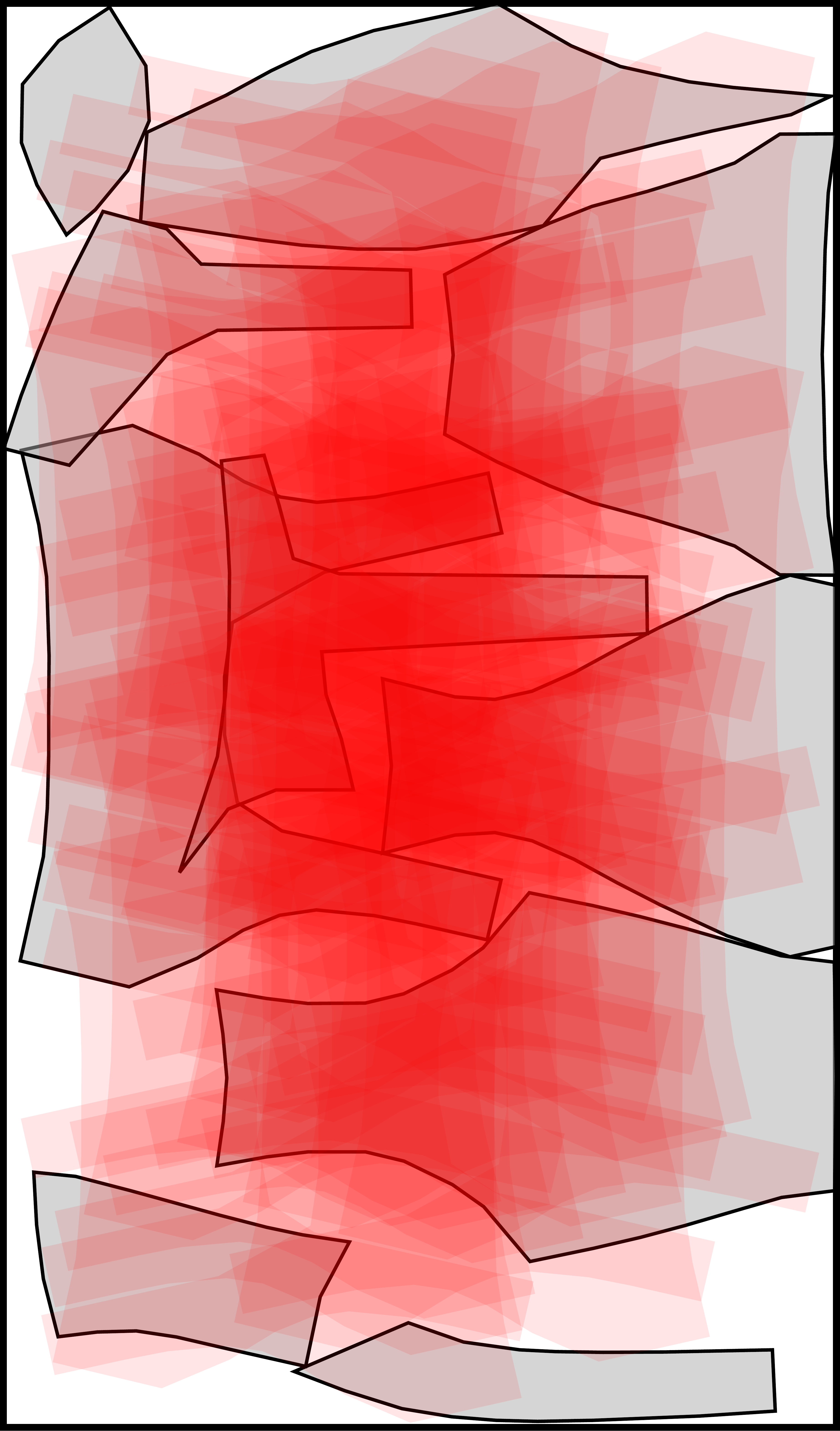}
        \caption{Random samples in the strip: $T_{div}$}
        \label{fig:sampling_bin}
    \end{subfigure}
    \begin{subfigure}[t]{.19\textwidth}
        \centering
        \includegraphics[height=\subfigheight]{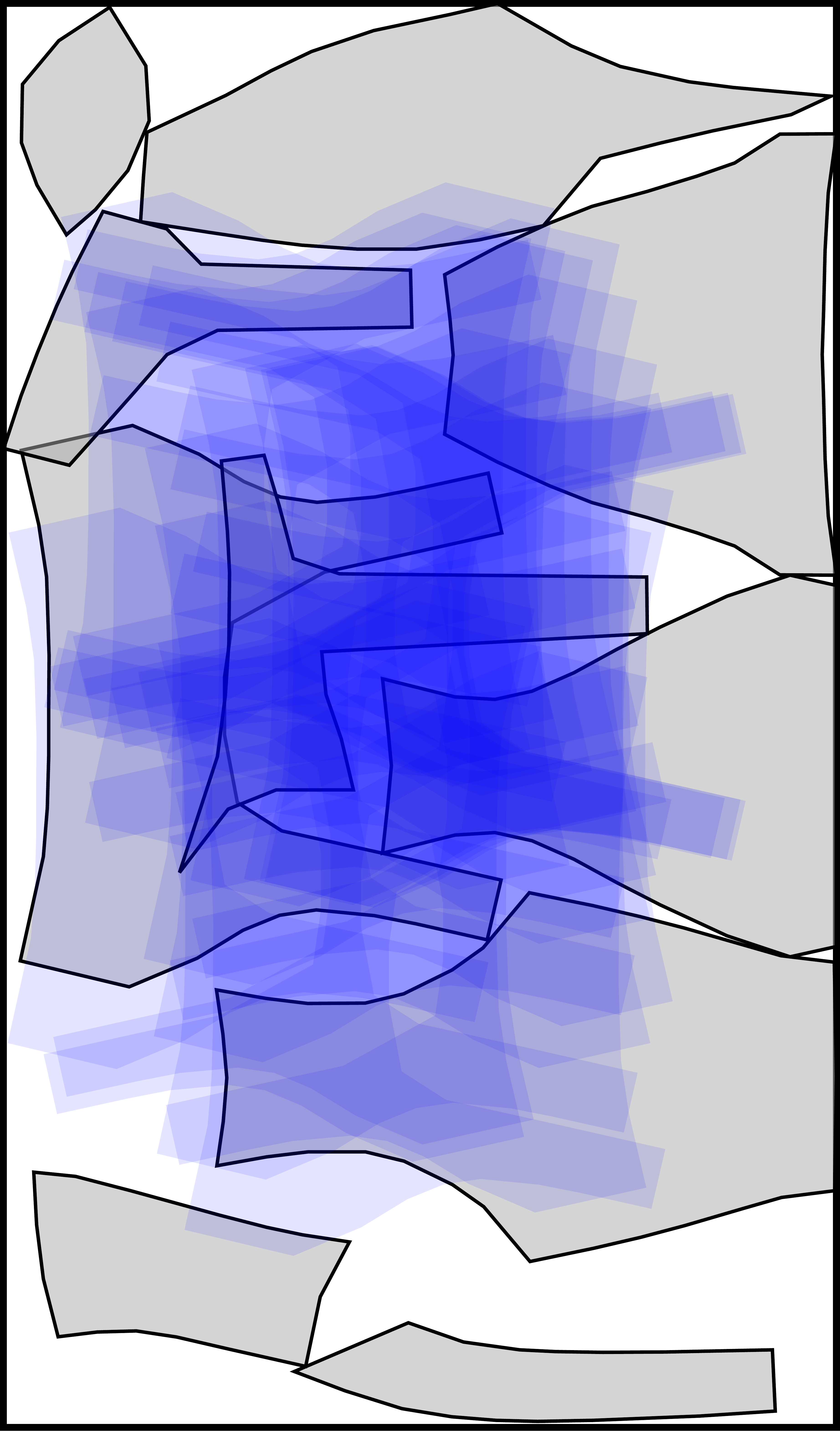}
        \caption{Random samples near the current position: $T_{foc}$}
        \label{fig:sampling_loc}
    \end{subfigure}
    \begin{subfigure}[t]{.19\textwidth}
        \centering
        \includegraphics[height=\subfigheight]{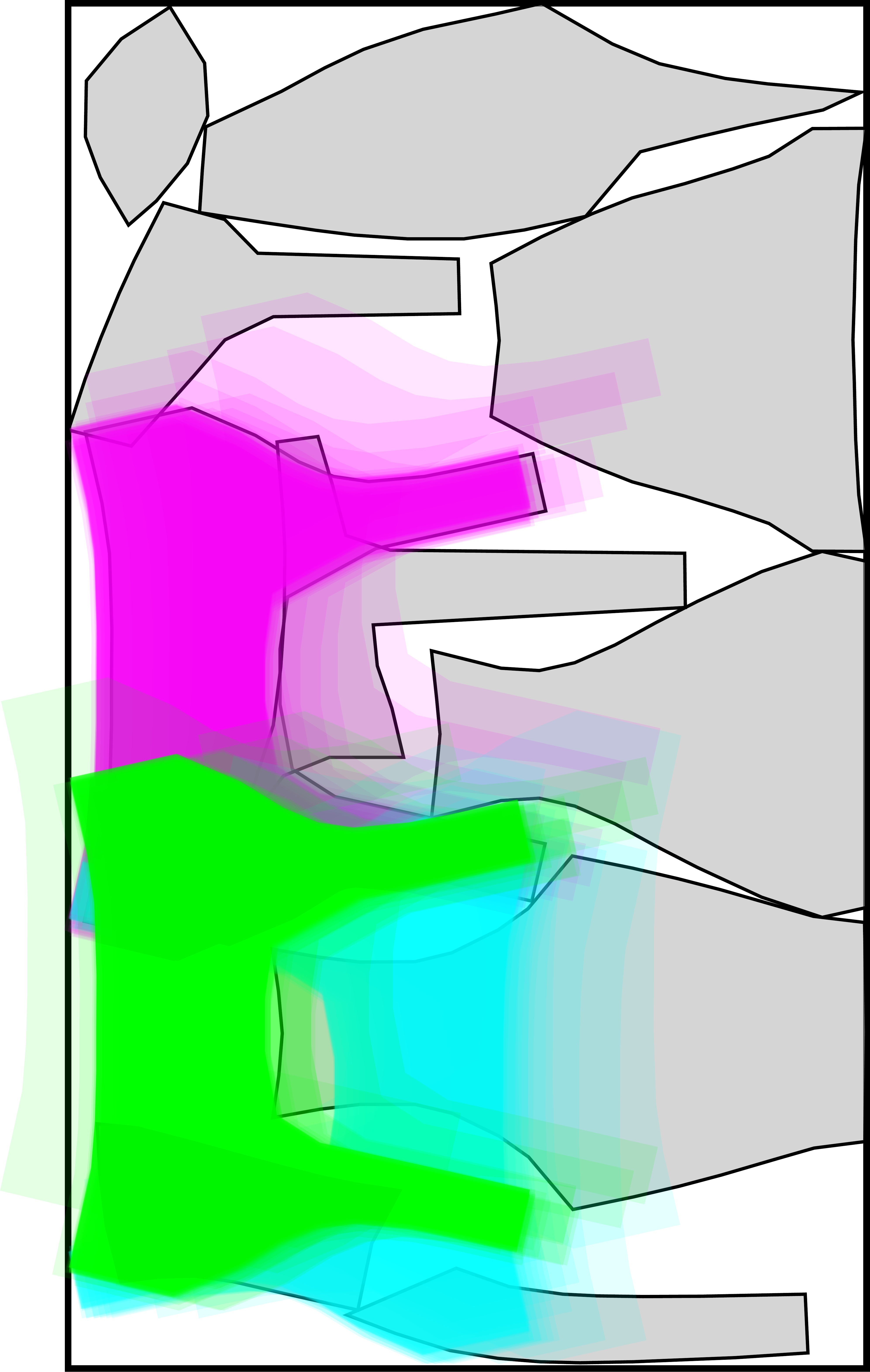}
        \caption{Refinement \\ of 3 samples}
        \label{fig:sampling_cd}
    \end{subfigure}
    \begin{subfigure}[t]{.19\textwidth}
        \centering
        \includegraphics[height=\subfigheight]{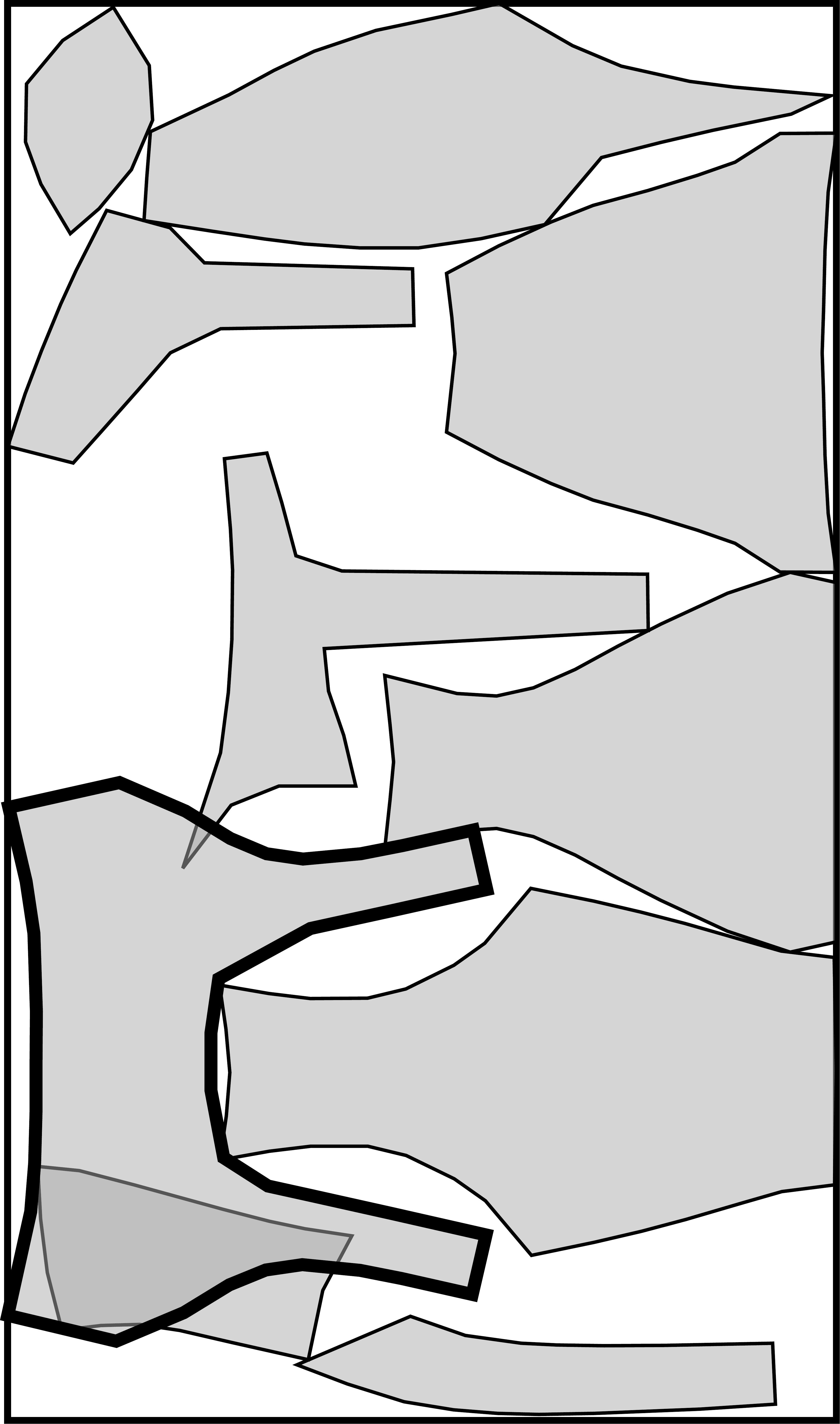}
        \caption{After}
        \label{fig:sampling_after}
    \end{subfigure}
    \caption{Searching for a new position for a single item (in bold) using the \texttt{perform\_sampling} procedure.}
    \label{fig:sampling}
\end{figure}

Figure \ref{fig:sampling} illustrates the complete process of searching for a new position for the item highlighted with a thick border in Figure \ref{fig:sampling_before}.
First, we generate a number of samples uniformly at random within the container -- the red overlays in Figure \ref{fig:sampling_bin} -- and collect them in a set $T_{div}$.
Next, we perform a number of samples uniformly at random in the vicinity of the item's current position -- the blue overlays in Figure \ref{fig:sampling_loc} -- resulting in set $T_{foc}$.
This leads to a set of samples $T_{div} \cup T_{foc}$, which is both \emph{diverse} within the container and also \emph{focused} around the item's current position.
While this set may contain promising samples, it is unlikely that any of them will yield a high-quality placement given they are all randomly generated.

We therefore select some of the most promising samples $\in T_{div} \cup T_{foc}$ to be \emph{refined} into their respective local optima.
A sample is considered promising if it is among the best evaluated and its transformation is sufficiently different from the other samples already selected.
Figure \ref{fig:sampling_cd} visualizes this refinement process for three samples that are highlighted in lime, cyan and purple.
The refinement procedure is inspired by the \emph{adaptive coordinate descent} approach described by \citet{loshchilov2011adaptive} and greedily shifts sampling in the direction of improving evaluation.
Figure \ref{fig:sampling_after} shows the position eventually chosen to move the item to.

Provided that the cost of evaluating a sample is low enough, this approach is an effective way to find high-quality placements while remaining completely agnostic to the underlying implementation of $f(i,t) \mapsto e$.

\section{Solving the feasibility problem} \label{section:solve_feas}
We have introduced how to detect collisions, how to quantify those collisions, and how to move from one solution to a neighboring one by repositioning colliding items.
The final step in solving the feasibility problem is to combine all these components into a complete \textbf{search algorithm}, which gradually resolves collisions with the ultimate aim of reaching a feasible configuration of items.

\subsection{Guided local search} \label{section:guided_local_search}
How to search for a new position was explained in Algorithm \ref{alg:search_position}, but the function to evaluate samples was left abstract.
A straightforward way to evaluate a sample would be to aggregate the severity of all collisions that would occur if the item were placed at the sampled position, akin to Algorithm \ref{alg:eval_pair}.
However, such a \emph{static} evaluation function would cause the search to converge to a local optimum within a couple of iterations.
Approaches proposed by \citet{umetani2009solving} and \citet{sato2019raster} avoid this issue by relying on Guided Local Search (GLS) \citep{voudouris2010guided}:
a metaheuristic that \emph{dynamically} modifies the evaluation function to escape local optima and move towards promising regions of the search space.

\begin{algorithm}[ht]
    \caption{\texttt{evaluate\_sample}$(i,t)$}
    \begin{algorithmic}[1]
        \State $C \gets \Call{jagua-rs::collisions}{t(S_i)}$ 
        \For{$c \in C \setminus \{i\}$} \Comment{ignore item $i$ itself}
            \State $w_{ic} \gets \text{weight of item-pair } (i,c)$
            \State $e \gets e + w_{ic} \cdot \Call{quantify\_collision}{t(S_i),S_c}$ \Comment{Alg. \ref{alg:quantify_collision}}
        \EndFor
        \State \Return $e$
    \end{algorithmic}
    \label{alg:eval_sample}
\end{algorithm}

Algorithm \ref{alg:eval_sample} introduces how we evaluate a sample $(i, t)$.
First, the CDE is queried to retrieve the set of items $C$ that would be colliding with item $i$ if it were moved to the position defined by transformation $t$.
These prospective collisions are quantified, multiplied by the corresponding item-pair weight $w_{ic}$, and summed up to a total evaluation value $e$.
These dynamic \textbf{item-pair weights} are the mechanism through which GLS modifies the evaluation function, thereby influencing the search process.

\begin{algorithm}[ht]
    \caption{\texttt{update\_weights}}
    \begin{algorithmic}[1]
        \State $e_\text{max} \gets$ max$\{$\nameref{alg:eval_pair}$ : a,b \in I,  a \neq b\}$  \Comment{Alg. \ref{alg:eval_pair}}
        \For{$a,b \in I: a \neq b$} \Comment{all pairs of items}
            \State $e \gets$ \nameref{alg:eval_pair} \Comment{Alg. \ref{alg:eval_pair}}
            \If{$e > 0$}
                \State $m \gets \texttt{M}_l + (\texttt{M}_u - \texttt{M}_l) \cdot (e / e_\text{max})$ \Comment{Table \ref{tab:params}}
            \Else
                \State $m \gets \texttt{M}_d$ \Comment{Table \ref{tab:params}}
            \EndIf
            \State $w_{ab} \gets$ max$\{1, w_{ab} \cdot m\}$
        \EndFor
    \end{algorithmic}
    \label{alg:iterate_weights}
\end{algorithm}

All item-pair weights are initialized to 1 and every time Algorithm \ref{alg:iterate_weights} is called, each weight is updated by multiplying it by its corresponding factor $m$.
If the pair of items is colliding, $m$ represents a linear mapping $\in \;\mathclose] \texttt{M}_l , \texttt{M}_u \mathclose]$ of the severity of their collision $e$ relative to the most severe collision $e_{\text{max}}$.
Item pairs with more severe collisions will thus have their weight rise faster than those with less severe collisions.
If a pair is not in collision, their weight is multiplied by $\texttt{M}_d$ ($<1$) and eventually decays back to 1 over time.
Parameters $\texttt{M}_l$, $\texttt{M}_u$ and $\texttt{M}_d$ are defined in Table \ref{tab:params}.

\subsection{Separation Procedure} \label{section:separation}

Simply looping \nameref{alg:move_items} (Algorithm \ref{alg:move_items}) and \nameref{alg:iterate_weights} (Algorithm \ref{alg:iterate_weights}) would technically suffice to traverse the solution space, but this approach lacks any mechanism to backtrack to previous solutions.
A more sophisticated approach that incorporates this ability should improve the effectiveness of the search algorithm.

Let us begin by defining the global objective function to evaluate an entire solution:
\begin{equation}
    z = \sum_{\substack{a, b \in I \\ a \ne b}} \text{\nameref{alg:eval_pair}}
    \label{eq:solution_loss}
\end{equation}
This function is an expression of the total severity of all collisions in the solution, computed as the sum of the evaluation of every item pair (Algorithm \ref{alg:eval_pair}).
Whenever $z = 0$, all collisions have been resolved and a feasible solution has been reached.

\begin{algorithm}[htbp]
    \caption{\texttt{separate}($k_\text{max}, n_\text{max}$)}
    \begin{algorithmic}[1]
        \State $s^* \gets$ current solution
        \SemiCol $z^* \gets \text{Expr. \ref{eq:solution_loss}}$
        \SemiCol $k \gets 0$
        \State $w_{ab} \gets 1$ for all item pairs $a,b \in I$ \Comment{GLS weights}
        \While{$k < k_\text{max}$ \textbf{and} $z^* > 0$}
            \State restore to ${s^*}$ 
            \SemiCol $s_\text{init} \gets s^*$
            \SemiCol $n \gets 0$ 
            \While{$n < n_\text{max}$ \textbf{and} $z^* > 0$}
            \Comment{an attempt}
                \State \nameref{alg:move_items_multi}
                \Comment{Alg. \ref{alg:move_items_multi}}
                \State \nameref{alg:iterate_weights}
                \Comment{Alg. \ref{alg:iterate_weights}}
                \State $z \gets \text{Expr. \ref{eq:solution_loss}}$
                \SemiCol $n \gets n + 1$
                \If{$z < z^*$} \Comment{improvement found}
                    \State $s^* \gets$ current solution
                    \SemiCol $z^* \gets z$
                    \State $n \gets 0$                    
                \EndIf
            \EndWhile
            \State $k \gets k + 1$ \Comment{add a strike}
            \If{$s^* \neq s_\text{init}$}
            \State $k \gets 0$ \Comment{reset strikes}
            \EndIf
        \EndWhile
        \State \Return $s^*$
    \end{algorithmic}
    \label{alg:separate}
\end{algorithm}

Algorithm \ref{alg:separate} describes the complete procedure to solve the feasibility problem introduced in Section \ref{section:feasibility_problem}.
The separation procedure represents a local search heuristic that starts from an infeasible situation with colliding items and attempts to resolve all of its collisions.
At its core, the algorithm continually repositions items, updates the item-pair weights and replaces the incumbent solution $s^*$ whenever improvements are found with respect the objective function in Expression \ref{eq:solution_loss}.
Whenever feasibility is reached, the procedure stops and immediately returns the solution.
Once a certain number of iterations without improvements $n_\text{max}$ is reached, we conclude one \emph{attempt}.
If the incumbent solution was never updated during this attempt, a \emph{strike} is added.
Conversely, if the attempt improved $s^*$, the strike counter $k$ is reset to 0.
As long as the number of strikes is below a certain threshold $k_\text{max}$, the procedure restores the incumbent solution $s^*$ and starts a new attempt, continuing its quest for a feasible solution.
Once the number of strikes exceeds $k_\text{max}$, the procedure halts the search and returns $s^*$, which will unfortunately still be infeasible.

The behavior we intend to generate through this procedure can be described as follows.
GLS ensures that item pairs colliding across consecutive iterations quickly reach prohibitively large weights, especially pairs that are involved in the most severe collisions.
By contrast, when a pair of items manages to resolve its collision, the weight slowly decays back down to 1.
When a collision occurs in a region with low GLS weights, the items involved will often `transfer' the collision to neighboring pairs.
Observing this process over multiple iterations reveals that items effectively \emph{push} and \emph{pull} each other, constantly `fighting' for their own space in the container.
When the search inevitably stalls, some items start to accumulate significant weights with all their neighbors.
This will cause them to \emph{jump} away to an entirely different position in the container, effectively escaping the local optimum.
Readers who are curious to see this behavior in action are referred to the repository (Section \ref{section:implementation}), which contains a simple visualizer.

\begin{figure}[ht]
    \centering
    \includegraphics[width=0.7\textwidth]{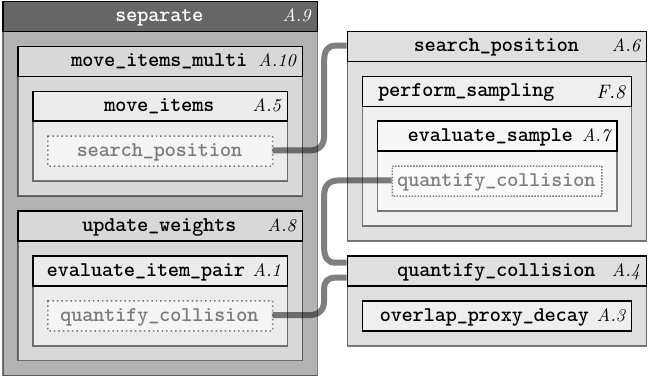}
    \caption{The \texttt{separate} procedure and all its underlying components presented in a hierarchical overview. Every component references its corresponding algorithm (\textit{A.$*$}) or figure (\textit{F.$*$}).}
    \label{fig:alg_overview}
\end{figure}

Figure \ref{fig:alg_overview} provides an overview of the \texttt{separate} procedure, where each of the previously introduced components are hierarchically related with one another.

\subsection{Balancing exploration and exploitation}\label{section:move_items_multi}

The random order in which Algorithm \ref{alg:move_items} repositions colliding items ensures that the search algorithm can explore a wide range of neighboring solutions.
However, in many cases, this random order might be far from ideal.
In order to strike a better balance between exploration and exploitation, Algorithm \ref{alg:move_items_multi} introduces an enhanced approach to move from one solution to a neighboring one.

\begin{algorithm}[htbp]
    \caption{\texttt{move\_items\_multi}}
    \begin{algorithmic}[1]
        \State $s_i \gets$ current solution \SemiCol $z^* \gets \infty$
        \For{\texttt{N\_WORKERS} times} \Comment{parallelized}      
            \State\nameref{alg:move_items} \Comment{Alg. \ref{alg:move_items}}
            \State $z \gets \text{Expr. {\ref{eq:solution_loss}}}$
            \If{$z < z^*$}
                \State $s^* \gets$ current solution
                \SemiCol $z^* \gets z$
            \EndIf
            \State restore to $s_i$
        \EndFor
        \State restore to $s^*$
    \end{algorithmic}
    \label{alg:move_items_multi}
\end{algorithm}

Algorithm \ref{alg:move_items_multi} essentially executes Algorithm \ref{alg:move_items} multiple times with different random orders of items.
From these multiple neighboring solutions, the best one (Expression \ref{eq:solution_loss}) is chosen.
This significantly increases the chances of finding improvement while preserving the diversity of the search process.
This enhanced version is employed in the separation procedure (Algorithm \ref{alg:separate}) to solve the feasibility problem.
Although presented here sequentially, Algorithm \ref{alg:move_items_multi} is parallelized across a pool of \texttt{N\_WORKERS} threads in our implementation.

\section{Solving the 2D irregular strip packing problem} \label{section:spp_heuristic}
Algorithm \ref{alg:separate}, which targets the feasibility problem introduced in Section \ref{section:feasibility_problem}, can now be integrated into a complete heuristic to solve the actual 2DISPP.
The complete heuristic we propose is introduced in Algorithm \ref{alg:solve} and is divided into two distinct phases.
The \textbf{exploration} phase first provides a lot of \emph{freedom} to enable investigation of a wide range of solutions.
Next, the \textbf{compression} phase attempts to shrink the strip as much as possible while remaining \emph{constrained} to the general configuration of items found at the end of the exploration phase.

\begin{algorithm}[htbp]
    \begin{algorithmic}[1]
        \State $s^* \gets$ \nameref{alg:explore} \Comment{Alg. \ref{alg:explore}}
        \State $s^* \gets$ \nameref{alg:compress} \Comment{Alg. \ref{alg:compress}}
        \State \Return $s^*$
    \end{algorithmic}
\caption{\texttt{solve\_ispp}}
\label{alg:solve}
\end{algorithm} 

\subsection{Exploration phase} \label{section:exploration_phase}

\begin{algorithm}[ht]
    \begin{algorithmic}[1]
        \State $s^* \gets$ construct initial solution and shrink strip by $\texttt{R}_{x}$
        \While{$\texttt{TL}_{x}$ not reached}
            \State $s \gets \Call{separate}{\texttt{K}_x,\texttt{N}_x}$ \Comment{Alg. \ref{alg:separate}}
            \If{$s$ \textbf{is} feasible}
                \State $s^* \gets s$
                \State shrink strip by $\texttt{R}_{x}$
                \State $\mathcal{S} \gets \emptyset $    
            \Else
                \State $\mathcal{S} \gets \mathcal{S} \cup \{s\}$ \Comment{pool of infeasible solutions}
                \State $\hat{s} \gets$ select from $\mathcal{S}$
                \State $s' \gets$ disrupt $\hat{s}$ by swapping two large items
                \State restore to $s'$   
            \EndIf
        \EndWhile
        \State \Return $s^*$
    \end{algorithmic}
    \caption{\texttt{explore}}
    \label{alg:explore}
\end{algorithm}

Algorithm \ref{alg:explore} describes the exploration phase, which starts from an initial solution created by a simple bottom-left-fill heuristic.
This constructive heuristic will provide a feasible solution, after which the strip is shrunk by a factor $\texttt{R}_{x}$.
Each time the strip is shrunk, all items have to be contained within these new boundaries.
This is achieved by selecting a certain vertical axis within the strip and shifting items positioned right of this axis to the left, most likely creating multiple minor collisions.

The separation procedure (Algorithm \ref{alg:separate}) attempts to convert this infeasible solution into a feasible one.
If successful, the returned solution $s$ replaces the incumbent solution $s^*$, after which the strip is once again shrunk.
If unsuccessful, $s$ represents an infeasible local optimum that is added to a pool of solutions $\mathcal{S}$.
For the next separation attempt, a random solution $\hat{s}$ is chosen from $\mathcal{S}$, with those closer to feasibility (Expression \ref{eq:solution_loss}) having a greater chance of being selected.
A new starting solution $s'$ is then created by swapping two large items from $\hat{s}$.
This ensures $s'$ inherits many high-quality placements from $\hat{s}$ while, at the same time, being sufficiently disrupted so it can escape from the local optimum.
This procedure continues until time limit $\texttt{TL}_x$ is reached, at which point the exploration phase returns the best feasible solution found $s^*$.

\subsection{Compression phase} \label{section:compression_phase}

\begin{algorithm}[htbp]
    \caption{\texttt{compress}($s^*$)}
    \begin{algorithmic}[1]
        \While {$\texttt{TL}_{c}$ not reached}
            \State $\tau \gets \text{elapsed time in phase}$
            \State $r \gets \texttt{R}_{c}^s + (\texttt{R}_{c}^e - \texttt{R}_{c}^s) \cdot (\tau / \texttt{TL}_c)$ \Comment{Table \ref{tab:params}}
            \State restore to $s^*$ and shrink strip by $r$
            \State $s' \gets \Call{separate}{\texttt{K}_c,\texttt{N}_c}$ \Comment{Alg. \ref{alg:separate}}
            \If{$s'$ \textbf{is} feasible} 
                \State $s^* \gets s'$
            \EndIf
        \EndWhile
        \State \Return $s^*$
    \end{algorithmic}
    \label{alg:compress}
\end{algorithm}

Algorithm \ref{alg:compress} describes the compression phase, which starts from the best feasible solution $s^*$ the exploration phase generated.
The procedure iteratively (i) restores to the incumbent solution $s^*$, (ii) shrinks the bin by ratio $r$ and (iii) attempts to separate.
If successful, the incumbent solution is updated.
If unsuccessful, the incumbent solution is restored and the process reattempted.

Shrink ratio $r$ is much smaller than during the exploration phase and decays linearly from $\texttt{R}_{c}^s$ to $\texttt{R}_{c}^e$ based on the phase's remaining time limit $\texttt{TL}_c$.
This means that the separate procedure only needs to resolve (progressively) less severe collisions, which it is often able to do without making major changes to the configuration of items.
Another important difference with the exploration phase is that, before each new separation attempt, the best feasible solution $s^*$ is always restored and the strip is shrunk again.
This causes the search process' freedom to be significantly more restricted than during the exploration phase.
The compression phase is squarely focused on finding out how much further the strip can be shrunk without deviating too far from the configuration of items provided by the exploration phase.

\section{Implementation} \label{section:implementation}
Our implementation of the proposed 2D nesting algorithm is written in Rust and called \texttt{sparrow}.
It is publicly available at: \url{https://github.com/JeroenGar/sparrow}.
The codebase closely follows the structure of this paper, with all introduced algorithms being referenced directly in the source code.
This makes it straightforward to understand how exactly \texttt{sparrow} maps to the algorithms presented here.
Throughout this project, we have strived for maximum computational performance, robustness and leanness of both the algorithm and its implementation.

The result of these priorities is an implementation that is relatively insensitive to the complexity and the number of shapes involved, making \texttt{sparrow} suitable as either a practical tool to tackle real-world problems or a basis for future research.
The contribution of each individual algorithmic component to the overall performance of \texttt{sparrow} was carefully evaluated and those without any significant contribution were eliminated.
As a result, the components presented in this paper represent an irreducible set required to achieve the performance demonstrated in Section \ref{section:experiments}.

It is important to note that, in contrast to some other approaches, \texttt{sparrow} will never produce a feasible solution where items are `touching' the container or each other.
Instead, items will always be completely separated, if only by a minuscule distance. 
Exact fits -- where moving the position of an item by even the smallest representable distance causes a collision -- cannot be produced by \texttt{sparrow}.

There are two reasons for this.
First, whenever two entities are extremely close to each other, \texttt{jagua-rs} will err on the side of caution and report a collision.
The CDE was deliberately designed this way in order to avoid numerical instability due to floating point arithmetic.
Second, when searching new positions for an item, \texttt{sparrow} relies on a sampling approach (Section \ref{section:continuous_search_space}).
While the sampling is configured to be very precise, it cannot target one specific coordinate.

At present, this limitation concerning exact fits only surfaces in artificial academic instances featuring simple shapes and vertices with integer coordinates.
In fact, real-world applications often require a minimum separation distance between items to account for manufacturing tolerances or cutting tool widths. This is a feature that \texttt{jagua-rs} already supports natively.

\section{Computational experiments} \label{section:experiments}
In this section we will evaluate the performance of \texttt{sparrow} by conducting a series of computational experiments on a set of academic benchmark instances.

\subsection{Experimental setup} \label{section:experimental_setup}
All experiments were performed on a machine equipped with an AMD Ryzen{\texttrademark} 9 7950X CPU. 
System memory is irrelevant as the entire program easily fits within the 64 MB of L3 cache of the CPU. 
Unless specified otherwise, each experiment was run for 20 minutes, with 3 threads active, and configured with the parameters presented in Table \ref{tab:params}.

\begin{table}[htbp]
\newcommand{\parcomment}[1]{\textcolor{gray}{\triangleright \textit{ #1}}}
\[
    \begin{array}{r @{\ } c @{\ } l @{\quad} l}
        \texttt{R}_\varepsilon & \gets & 1\% & \parcomment{Alg. \ref{alg:overlap_area_proxy_decay} -- diameter ratio to compute \(\varepsilon\)} \\
        (\texttt{M}_u, \texttt{M}_l, \texttt{M}_d) & \gets & (2.0, 1.2, 0.95) & \parcomment{Alg. \ref{alg:iterate_weights} -- upper, lower and decay weight multipliers} \\
        \texttt{N\_WORKERS} & \gets & 3 & \parcomment{Alg. \ref{alg:move_items_multi} -- how many times \texttt{move\_items} is executed} \\
        (\texttt{K}_x, \texttt{N}_x) & \gets & (3, 200) & \parcomment{Alg. \ref{alg:explore} -- separation parameters for exploration} \\
        (\texttt{K}_c, \texttt{N}_c) & \gets & (5, 100) & \parcomment{Alg. \ref{alg:compress} -- separation parameters for compression} \\
        \texttt{R}_x & \gets & 0.1\% & \parcomment{Alg. \ref{alg:explore} -- exploration shrink ratio} \\
        (\texttt{R}_c^s, \texttt{R}_c^e) & \gets & (0.05\%, 0.001\%) & \parcomment{Alg. \ref{alg:compress} -- compression shrink ratio range} \\
        (\texttt{TL}_x, \texttt{TL}_c) & \gets & (0.8, 0.2) \cdot 20' & \parcomment{Alg. \ref{alg:explore},\ref{alg:compress} -- time limits} \\
    \end{array}
\]
\caption{Configuration of \texttt{sparrow}'s parameters throughout the experiments.
The right column contains the relevant algorithm(s) and a short description for every entry.}
\label{tab:params}
\end{table}

We used a combination of manual tuning and hyperparameter optimization \citep{bergstra2013making} to identify a suitable set of parameters for the algorithm.
These parameters were optimized for the maximum expected performance across all instances from the academic benchmark set when subject to a 20-minute time limit.
This single set of parameters was used for every instance in our experiments.

\subsection{Performance analysis} \label{section:performance_analysis}
To evaluate the performance of \texttt{sparrow} we will use academic benchmark instances from the ESICUP website\footnote{\url{https://www.euro-online.org/websites/esicup/data-sets}}.
We refer to Table 3 in \citet{sato2019raster} for an overview of these instances and their properties.

To provide a sense of the efficiency of the implementation and scale of operations, when using the aforementioned experimental setup, \texttt{sparrow} is able to evaluate samples (Algorithm~\ref{alg:eval_sample}) at a rate of approximately 2.4 million per second with \texttt{SWIM}, which represents the most complex instance in the academic benchmark set.

Throughout the experiments, we report the quality of a solution in terms of its packing density $\rho$. 
This is the ratio of the area of the items in the strip with respect to the area of the strip itself expressed as a percentage:
\begin{equation}
    \rho = \frac{100}{w \cdot l} \cdot \sum_{i \in I} \texttt{area}(S_i)
\end{equation}

\begin{figure}[!htbp]
    \centering
    \includegraphics[width=0.95\textwidth]{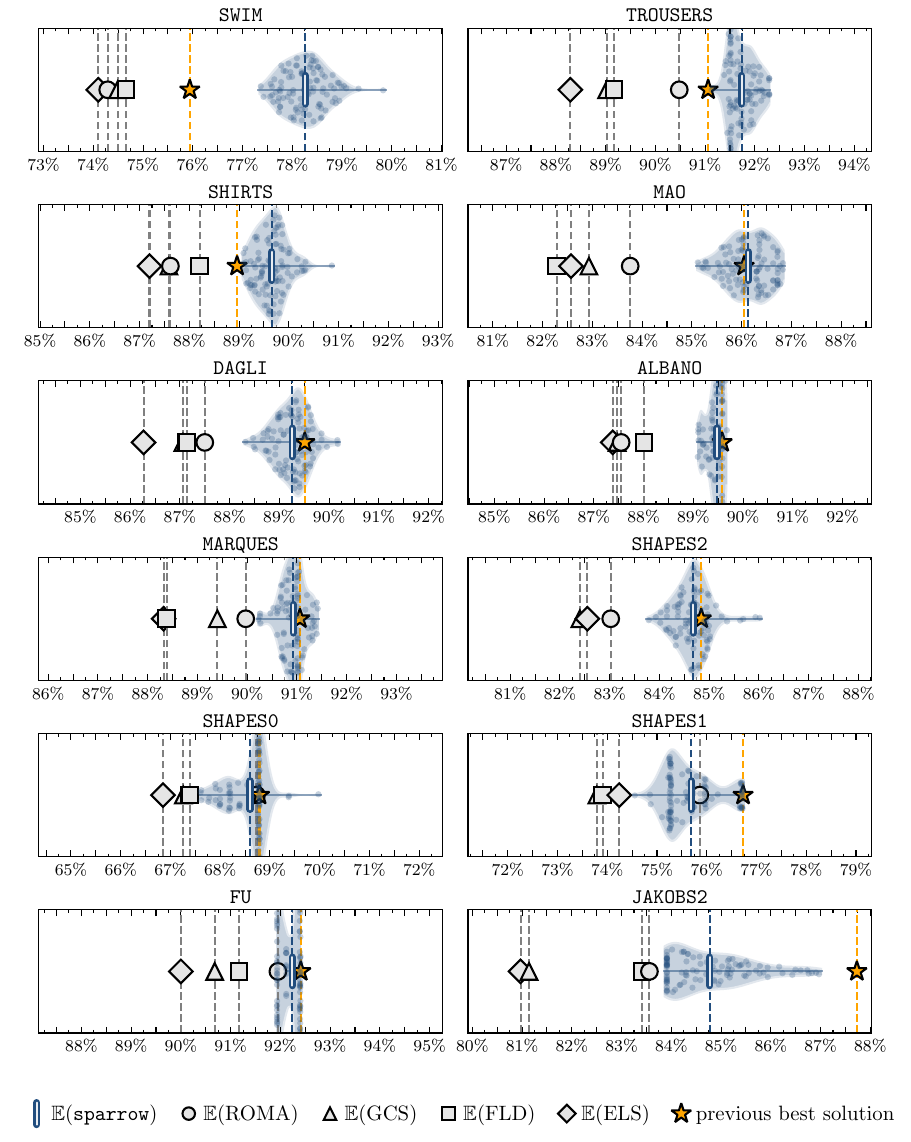}
    \caption{
        Distribution of \texttt{sparrow}'s solution quality $\rho$ across the academic benchmark instances.
        Each dot represents an individual solution from the experiments composing Table \ref{tab:comparison}.
        The expected solution qualities of ROMA - \cite{sato2019raster}, GCS - \cite{elkeran2013new}, FLD - \cite{wang2017flexible} and ELS - \cite{leung2012extended} algorithms are shown.
        A star marks the previous singular best-known solution for each instance.
        \texttt{JAKOBS1} is omitted since it virtually always produced the same density.
    }
    \label{fig:violin_plots}
\end{figure}

The expected quality $\mathbb{E}(\rho)$ and interquartile range of \texttt{sparrow}'s solutions in comparison to the academic state of the art algorithms for the 2DISPP and one open-source implementation are presented in Table \ref{tab:comparison}.
The distributions of the solutions from Table \ref{tab:comparison} are visualized as violin plots in Figure \ref{fig:violin_plots}. 
We performed 100 independent runs for each instance to ensure our reported expected values are statistically robust, with sufficiently narrow confidence intervals.
It is important to note that the width of the strip for some of the most simple instances -- \texttt{SHAPES0, SHAPES1, SHAPES2, FU, JAKOBS1 and JAKOBS2} -- was inflated by 0.01\%.
This minuscule inflation does not affect the possible feasible item configurations, but suffices to avoid the issue concerning exact fits (described in Section \ref{section:implementation}), which would otherwise occur in these artificial instances.
For academic purposes, \texttt{sparrow}'s best solution ever produced for each instance, alongside the previous best-known solution, is presented in Table~\ref{tab:best} and shown in Figure~\ref{fig:best_solutions}.
It is important to note that these `best' solutions, although conceived under the same experimental conditions and time limit, are not necessarily elements of the sets of 100 independent runs from which the expected densities in Table~\ref{tab:comparison} have been derived.

\begin{table}[!tbp]
   \centering
   \setlength{\tabcolsep}{4pt} 

\begin{tabular}{|l|cc|c|cccc|c|}
   \hline
& \multicolumn{3}{c|}{\texttt{sparrow}} & ROMA & GCS& FLD & ELS & PS \\
instance & {Q1$(\rho)$} & {Q3$(\rho)$} & { $\mathbb{E}$$(\rho)$} & { $\mathbb{E}$$(\rho)$} & { $\mathbb{E}$$(\rho)$} & { $\mathbb{E}$$(\rho)$} & { $\mathbb{E}$$(\rho)$} & {$\rho$} \\ \hline
\texttt{ALBANO} &89.39&89.58& \textbf{89.47} \scriptsize$\pm$0.03& 87.55 & 87.47 & 88.01 & 87.38 & 84.95 \\
\texttt{DAGLI} &89.01&89.50& \textbf{89.26} \scriptsize$\pm$0.07& 87.50 & 87.06 & 87.14 & 86.27 & 84.17 \\
\texttt{FU} &91.93&92.40& \textbf{92.24} \scriptsize$\pm$0.04& 91.95 & 90.68 & 91.17 & 90.00 & 90.39 \\
\texttt{JAKOBS1} &89.09&89.09& \textbf{89.09} \scriptsize$\pm$0.00& \textbf{89.09} & 88.90 & 88.96 & 88.35 & 81.67 \\
\texttt{JAKOBS2} &83.91&85.25& \textbf{84.77} \scriptsize$\pm$0.17& 83.56 & 81.14 & 83.41 & 80.97 & 80.42 \\
\texttt{MAO} &85.84&86.45& \textbf{86.14} \scriptsize$\pm$0.08& 83.76 & 82.93 & 82.28 & 82.57 & 75.94 \\
\texttt{MARQUES} &90.78&91.02& \textbf{90.93} \scriptsize$\pm$0.05& 89.97 & 89.40 & 88.38 & 88.32 & 85.48 \\
\texttt{SHAPES0} &68.38&68.79&68.60 \scriptsize$\pm$0.07& \textbf{68.73} & 67.26 & 67.39 & 66.85 & 66.50 \\
\texttt{SHAPES1} &75.26&75.97&75.69 \scriptsize$\pm$0.11& \textbf{75.86} & 73.79 & 73.91 & 74.24 & 72.55 \\
\texttt{SHAPES2} &84.44&84.88&84.68 \scriptsize$\pm$0.07& 83.02 & 82.40 & - & 82.55 & \textbf{85.49} \\
\texttt{SHIRTS} &89.43&89.84& \textbf{89.66} \scriptsize$\pm$0.06& 87.62 & 87.59 & 88.21 & 87.20 & 85.99 \\
\texttt{SWIM} &77.94&78.55& \textbf{78.26} \scriptsize $\pm$0.09& 74.29 & 74.49 & 74.66 & 74.10 & 71.44 \\
\texttt{TROUSERS} &91.51&91.93& \textbf{91.73} \scriptsize$\pm$0.05& 90.48 & 89.02 & 89.17 & 88.29 & 89.30 \\
\hline
\end{tabular}

\captionsetup{width=0.95\textwidth}
\caption{
   The expected solution quality $\mathbb{E}(\rho)$ on academic benchmarks obtained by \texttt{sparrow}, four algorithms from the literature: ROMA - \cite{sato2019raster}, GCS - \cite{elkeran2013new}, FLD - \cite{wang2017flexible}, ELS - \cite{leung2012extended} and one open-source implementation: PS - \cite{fontan}. 
   For \texttt{sparrow}, quartiles (Q1, Q3) and the margin of error ($\pm$) on $\mathbb{E}$, defined as the half-width of the 95\% confidence interval computed via non-parametric bootstrapping, are reported.
   The reported values for \texttt{sparrow} result from 100 independent runs, each with a time limit of 20 minutes. 
   Values for ROMA, GCS, FLD and ELS were taken from their respective publications.
   PS is a deterministic algorithm and was run once under the same conditions as \texttt{sparrow}.
   For each instance, the best $\mathbb{E}(\rho)$ is highlighted in bold.
}
\label{tab:comparison}
\end{table}

\newcommand{\trousersrecord}{92.62}
\newcommand{\shapeszerorecord}{69.98}
\newcommand{\swimrecord}{79.83}
\newcommand{\jakobsonerecord}{89.26}
\newcommand{\shirtsrecord}{90.92}

\begin{table}[!htbp]
    \centering
    \setlength{\tabcolsep}{4pt} 
  \begin{tabular}{|l|c|cc|}
     \hline
  & \texttt{sparrow} & \multicolumn{2}{c|}{previous best} \\
  instance & $\rho$ & $\rho$ & by \\ \hline
  \texttt{ALBANO} & \textbf{89.82} & 89.58 & \tiny{E} \\
  \texttt{DAGLI} & \textbf{90.17} & 89.51 & \tiny{E} \\
  \texttt{FU} & \textbf{92.41} & \textbf{92.41} & \tiny{E} \\
  \texttt{JAKOBS1} & \textbf{\jakobsonerecord} & 89.09 & \tiny{L/E/W/S} \\
  \texttt{JAKOBS2} & \textbf{87.73} & \textbf{87.73} & \tiny{E/S} \\
  \texttt{MAO} & \textbf{86.87} & 86.05 & \tiny{S} \\
  \texttt{MARQUES} & \textbf{92.02} & 91.02 & \tiny{S} \\
  \texttt{SHAPES0} & \textbf{\shapeszerorecord} & 68.79 & \tiny{E/W/S} \\
  \texttt{SHAPES1} & \textbf{76.73} & \textbf{76.73} & \tiny{E/S} \\
  \texttt{SHAPES2} & \textbf{86.23} & 84.84 & \tiny{E} \\
  \texttt{SHIRTS} & \textbf{\shirtsrecord} & 88.96 & \tiny{E/W} \\
  \texttt{SWIM} & \textbf{\swimrecord} & 75.94 & \tiny{E} \\
  \texttt{TROUSERS} & \textbf{\trousersrecord} & 91.06 & \tiny{S} \\
  \hline
  \end{tabular}
  \caption{The best solution densities ever produced by \texttt{sparrow} for instances in the benchmark set, accompanied by the previous best-known density and the algorithm that produced it:
  \textsc{l} - \cite{leung2012extended}.
  \textsc{e} - \cite{elkeran2013new},
  \textsc{w} - \cite{wang2017flexible}, 
  \textsc{s} - \cite{sato2019raster}.
  Best densities are highlighted in bold.
  Time limit is 20 minutes.
  }
\label{tab:best}
\end{table}

\begin{figure}[!htbp]
    \def\subfigheight{4.5cm}
    \begin{subfigure}[t]{0.30\textwidth}
        \includegraphics[height=\subfigheight]{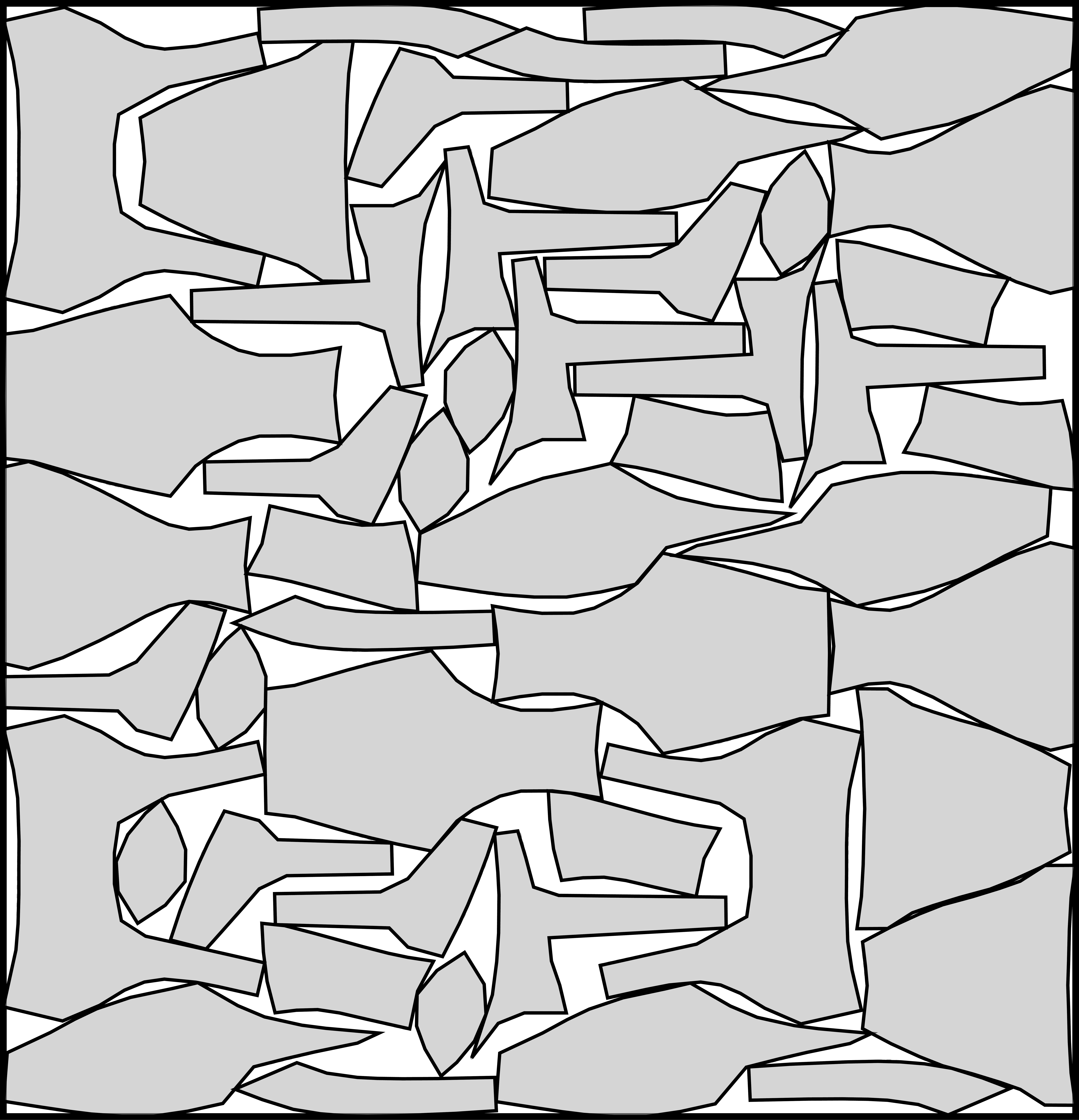}
        \caption{\texttt{SWIM} - \textbf{\swimrecord}\%}
    \end{subfigure}
    \hfill
    \begin{subfigure}[t]{0.45\textwidth}
        \centering
        \includegraphics[height=\subfigheight]{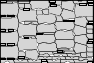}
        \caption{\texttt{SHIRTS} - \textbf{\shirtsrecord}\%}
    \end{subfigure}
    \hfill
    \begin{subfigure}[t]{0.22\textwidth}
        \hfill
        \includegraphics[height=\subfigheight]{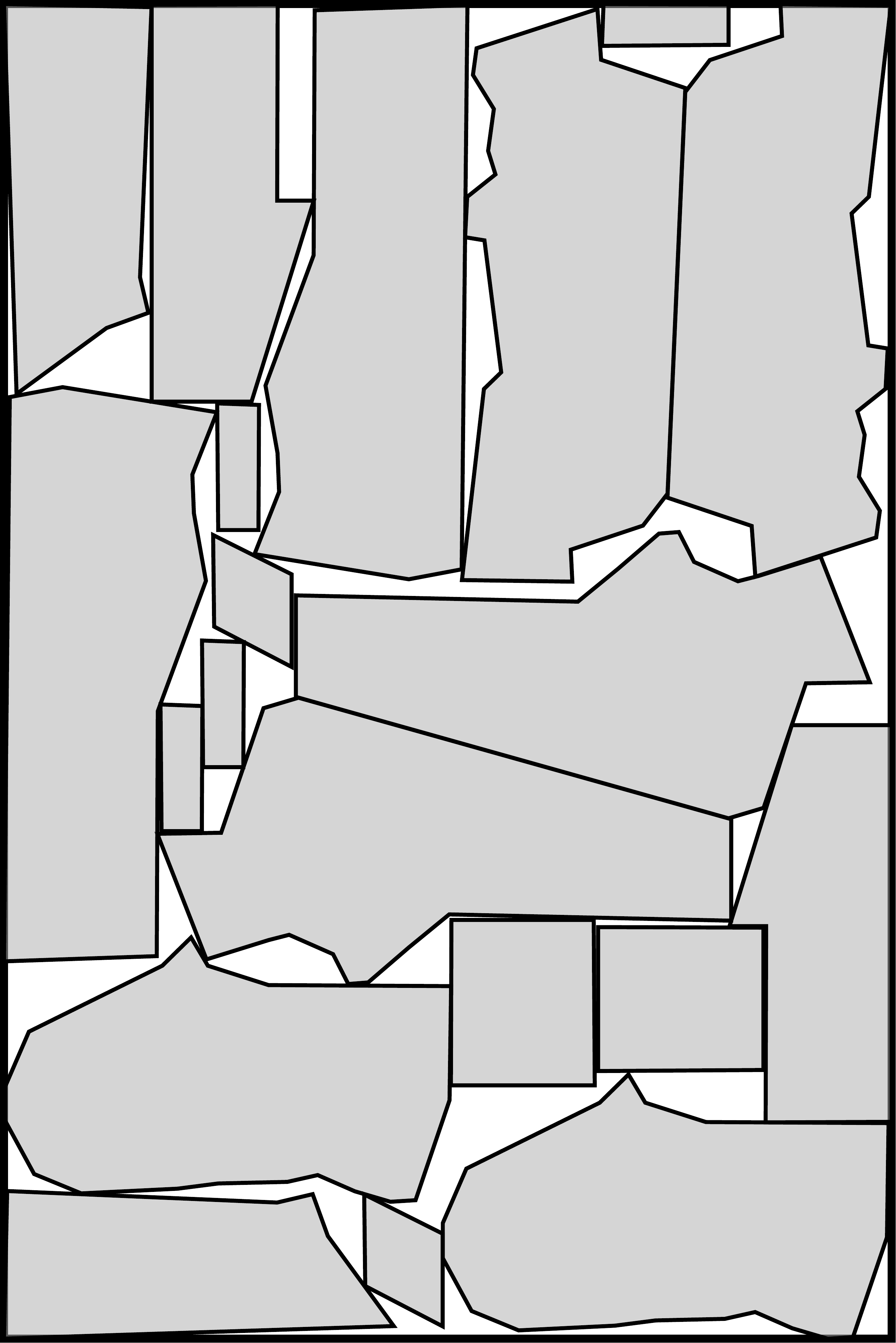}
        \caption{\texttt{MAO} - \textbf{86.87}\%}
    \end{subfigure}

    \vspace{.5em}
        
    \def\subfigheight{3.2cm}
    \begin{subfigure}[t]{0.43\textwidth}
        \includegraphics[height=\subfigheight]{final_best_albano_svg-raw.pdf}
        \caption{\texttt{ALBANO} - \textbf{89.82}\%}
    \end{subfigure}
    \hfill
    \begin{subfigure}[t]{0.36\textwidth}
        \centering
        \includegraphics[height=\subfigheight]{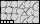}
        \caption{\texttt{SHAPES2} - \textbf{86.23}\%}
    \end{subfigure}
    \hfill
    \begin{subfigure}[t]{0.16\textwidth}
        \hfill
        \includegraphics[height=\subfigheight]{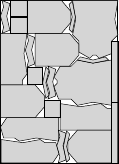}
        \caption{\texttt{MARQUES} \\ - \textbf{92.02}\%}
    \end{subfigure}
    
    \vspace{.5em}
    
    \def\subfigheight{3.8cm}
    \begin{subfigure}[t]{0.76\linewidth}
        \includegraphics[height=\subfigheight]{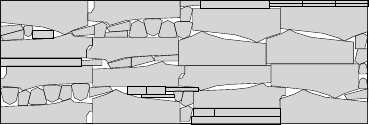}
        \caption{\texttt{TROUSERS} - \textbf{\trousersrecord}\%}
    \end{subfigure}
    \hfill
    \begin{subfigure}[t]{0.11\linewidth}
        \centering 
        \includegraphics[height=\subfigheight]{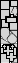}
        \caption{\texttt{JAKOBS1}\\ - \textbf{\jakobsonerecord}\%}
    \end{subfigure}
    \hfill
    \begin{subfigure}[t]{0.10\linewidth}
        \hfill
        \includegraphics[height=\subfigheight]{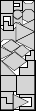}
        \caption{\texttt{JAKOBS2}\\ - 87.73\%}
    \end{subfigure}

    \vspace{.5em}

    \def\subfigheight{3.1cm}
    \begin{subfigure}[t]{0.2\linewidth}
        \includegraphics[height=\subfigheight]{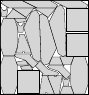}
        \caption{\texttt{DAGLI} - \textbf{90.17}\%}
    \end{subfigure}
    \hfill
    \begin{subfigure}[t]{0.29\linewidth}
        \centering
        \includegraphics[height=\subfigheight]{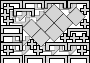}
        \caption{\texttt{SHAPES0} - \textbf{\shapeszerorecord}\%}
    \end{subfigure}
    \hfill
    \begin{subfigure}[t]{0.29\linewidth}
        \centering
        \includegraphics[height=\subfigheight]{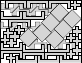}
        \caption{\texttt{SHAPES1} - 76.73\%}
    \end{subfigure}
    \hfill
    \begin{subfigure}[t]{0.17\linewidth}
        \hfill
        \includegraphics[height=\subfigheight]{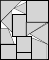}
        \caption{\texttt{FU} - 92.41\%}
    \end{subfigure}

    \caption{The best solutions obtained by \texttt{sparrow} (Table \ref{tab:best}) and their respective densities. Bold values indicate improved best-known solutions.}
    \label{fig:best_solutions}
\end{figure}

\texttt{sparrow} generally dominates in terms of expected solution quality, both on individual instances and across the entire benchmark set. 
Figure \ref{fig:violin_plots} reveals that for the three largest instances in terms of number of items and complexity of their shapes -- \texttt{SWIM}, \texttt{SHIRTS} and \texttt{TROUSERS} -- \texttt{sparrow}'s performance distribution lies entirely beyond the previous best-known singular solution.
Indeed, we were able to match or improve upon every previous best-known solution, in some cases by a significant margin.

The algorithms composing the academic state of the art report only two values per instance: the best and the expected density from a limited number of runs.
This limits our ability to provide detailed statistical or visual performance comparisons against them.
However, to support future in-depth comparisons against \texttt{sparrow}, we provide all solution data and a complete reproducibility guide in the repository.

In conclusion, \texttt{sparrow} establishes a new state of the art for the 2DISPP, consistently producing solutions whose quality, in most cases, significantly exceeds any previous academic or open-source results.

\subsection{Time-sensitivity analysis}
\label{section:time_sensitivity}
\begin{figure}[!htbp]
    \centering
    \includegraphics[]{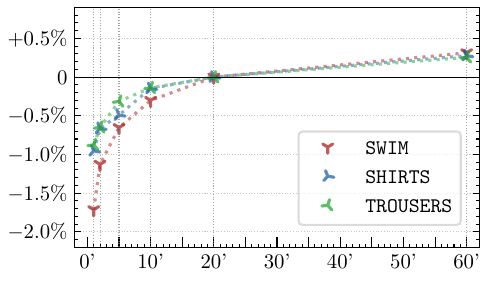}
    \caption{
        Scaling of the expected solution quality $\mathbb{E}(\rho)$ of \texttt{sparrow} for the three most challenging instances across different time limits, normalized for 20 minutes.
        Each data point is composed of 20 independent runs and no specific tuning was performed to account for the different time limits.
        }
    \label{fig:time-sensitivity}
\end{figure}

Different use cases call for varying computational budgets, and therefore it is important to understand how performance scales with time.
Figure \ref{fig:time-sensitivity} presents a time-sensitivity analysis of \texttt{sparrow}'s expected performance for the three most complex instances from the academic benchmark set, normalized to the 20-minute time limit from Table \ref{tab:comparison}.
No specific tuning was performed to account for the different time limits.
\texttt{sparrow} is able to quickly produce solutions within a couple of percentage points of the achievable quality after just 20 minutes. 
Beyond 20 minutes, the expected solution quality continues to improve, albeit at a diminishing rate.

\section{Real-world benchmarks \& future research directions} \label{section:new_instances}
Section \ref{section:performance_analysis} demonstrated that \texttt{sparrow} is capable of consistently producing high-quality solutions across the entire set of traditional academic benchmark instances.
This set, however, is dated and contains many instances that were constructed artificially, featuring simple polygonal shapes with vertices located at integer coordinates.
For \texttt{SWIM}, \texttt{TROUSERS} and \texttt{SHIRTS}, arguably the instances that most reflect real-world conditions, \texttt{sparrow} makes a significant leap forward.
However, these instances are few in number and only represent a single application domain: the garment industry.
We therefore take the opportunity to extend the set of academic benchmark instances
to better capture the breadth and intricacies of real-world applications.

We introduce ten new instances that span a variety of application domains and contain both homogeneous and heterogeneous sets of items, as well as narrow and wider strips.
Every item is represented as a simple polygon and two variants are provided for each instance: one which only allows discrete rotations in 90-degree increments and a second which allows continuous rotation.
The full dataset and experimental results are available in \texttt{sparrow}'s repository and in the ESICUP datasets\footnote{\url{https://github.com/ESICUP/datasets}}.

\begin{figure}[!htbp]
    \def\subfigheight{1.8cm}
    \begin{subfigure}[t]{0.295\textwidth}
        \includegraphics[height=\subfigheight]{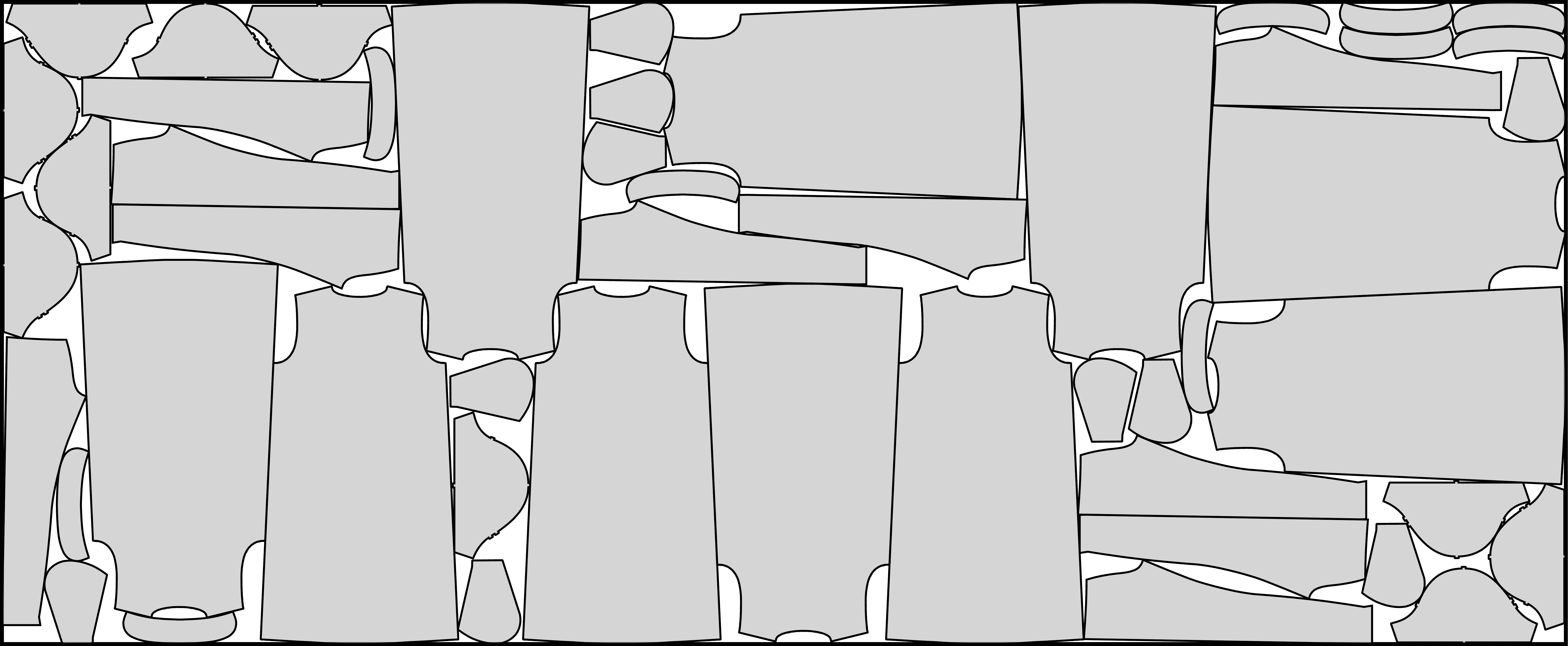}
        \caption{\texttt{GARDEYN0}}
    \end{subfigure}
    \hfill
    \begin{subfigure}[t]{0.12\textwidth}
        \centering
        \includegraphics[height=\subfigheight]{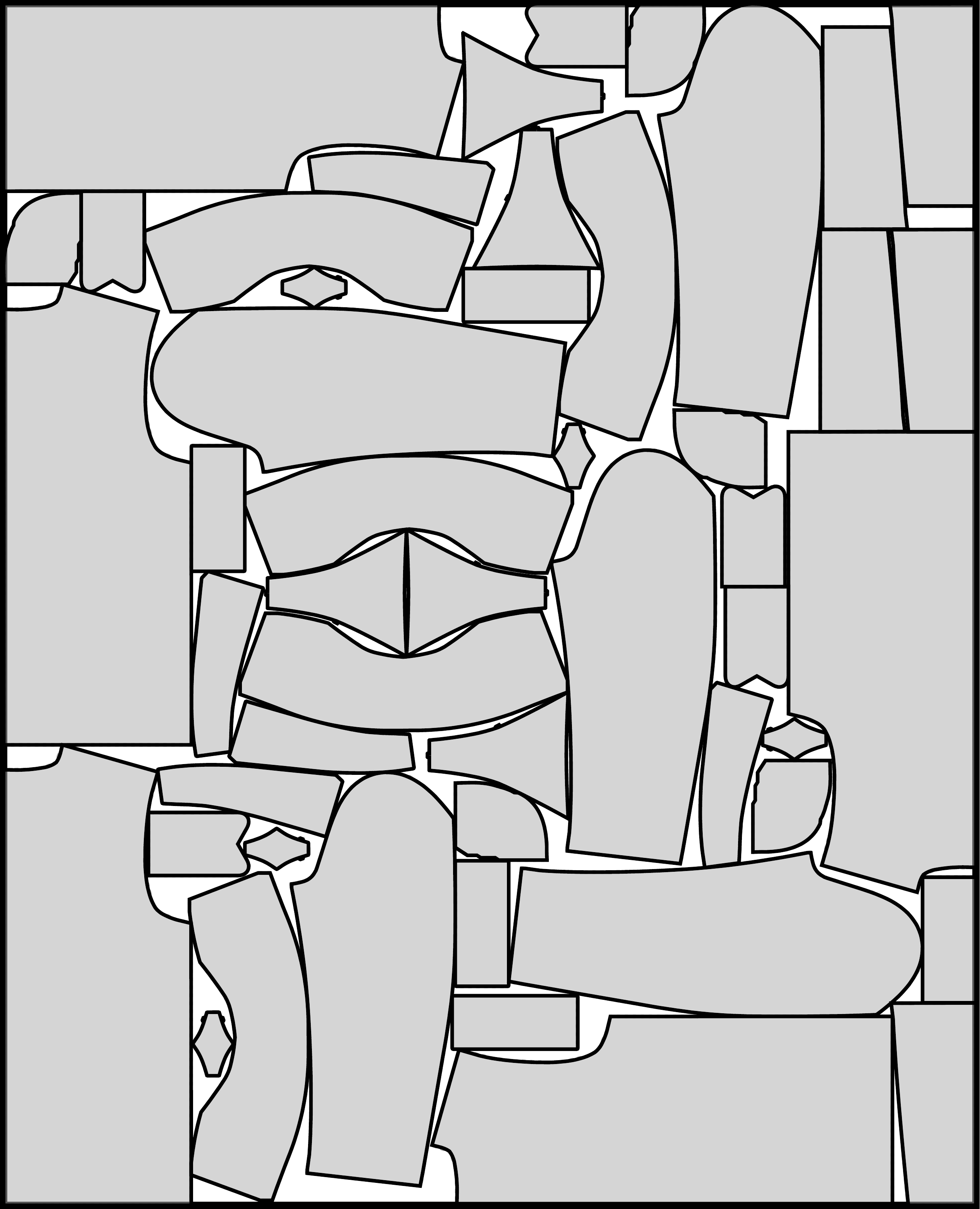}
        \caption{\texttt{GARDEYN1}}
    \end{subfigure}
    \hfill
    \begin{subfigure}[t]{0.57\textwidth}
        \hfill
        \includegraphics[height=\subfigheight]{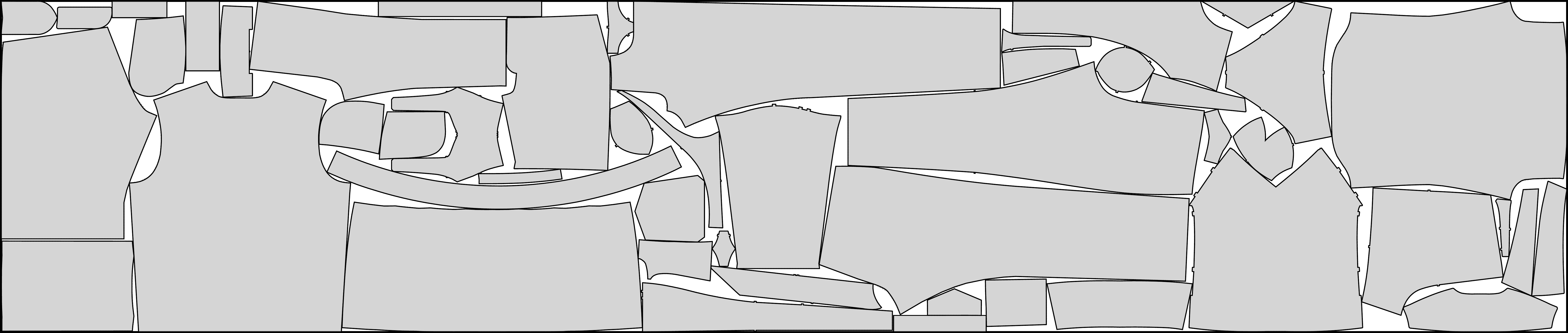}
        \caption{\texttt{GARDEYN2}}
    \end{subfigure}

    \vspace{.5em}
        
    \def\subfigheight{2.4cm}
    \begin{subfigure}[t]{0.48\textwidth}
        \includegraphics[height=\subfigheight]{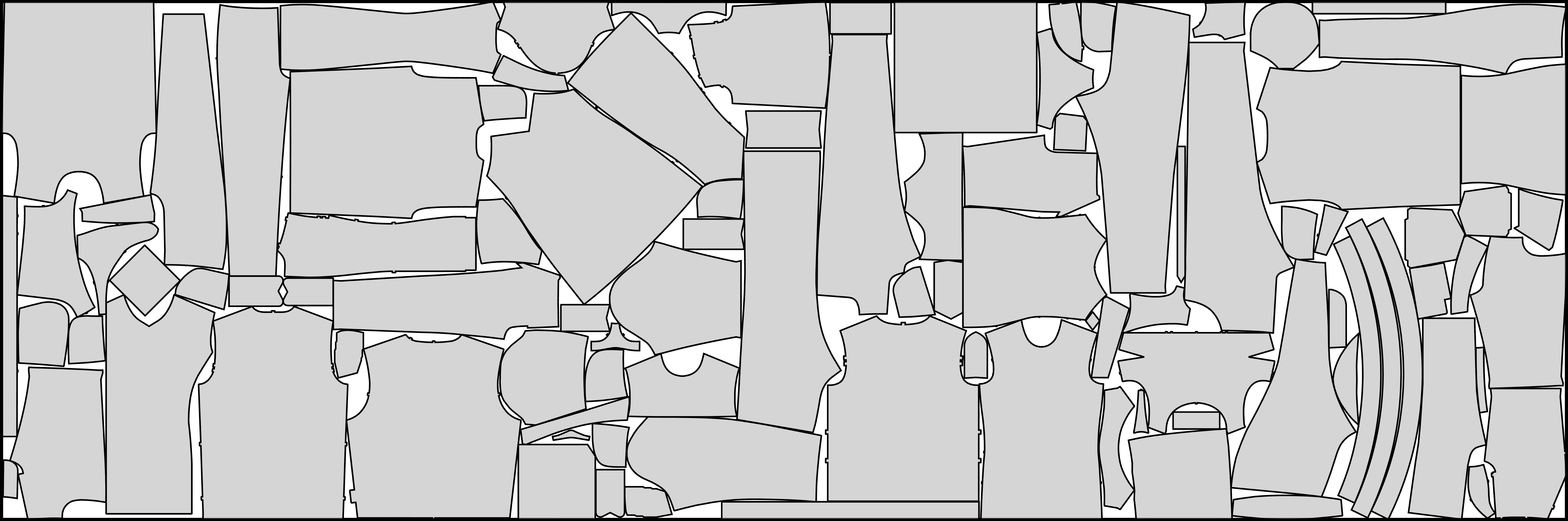}
        \caption{\texttt{GARDEYN3}}
    \end{subfigure}
    \hfill
    \begin{subfigure}[t]{0.51\textwidth}
        \hfill
        \includegraphics[height=\subfigheight]{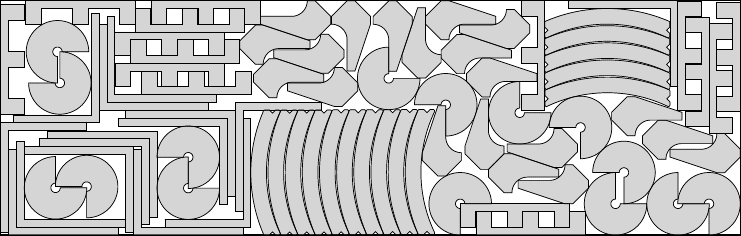}
        \caption{\texttt{GARDEYN4}}
    \end{subfigure}

    \vspace{.5em}

    \def\subfigheight{2.3cm}
    \begin{subfigure}[t]{0.33\textwidth}
        \includegraphics[height=\subfigheight]{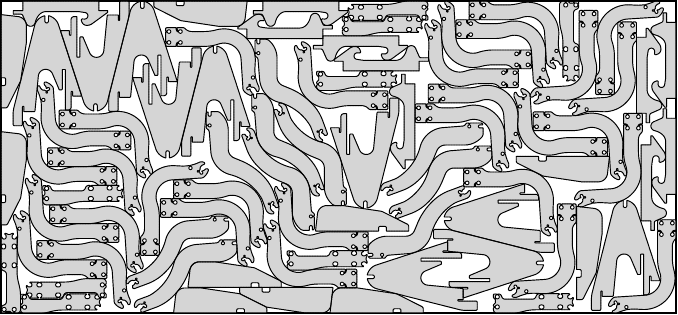}
        \caption{\texttt{GARDEYN5}}
    \end{subfigure}
    \hfill
    \begin{subfigure}[t]{0.65\linewidth}
        \hfill
        \includegraphics[height=\subfigheight]{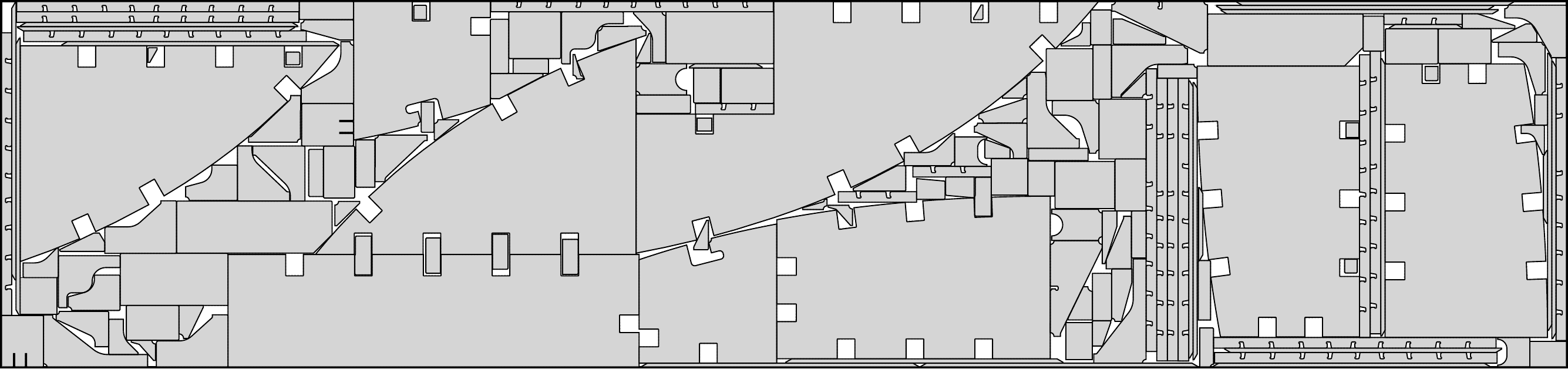}
        \caption{\texttt{GARDEYN6}}
    \end{subfigure}
    
    \vspace{.5em}

    \def\subfigheight{2cm}

    \begin{subfigure}[t]{0.44\linewidth}
        \includegraphics[height=\subfigheight]{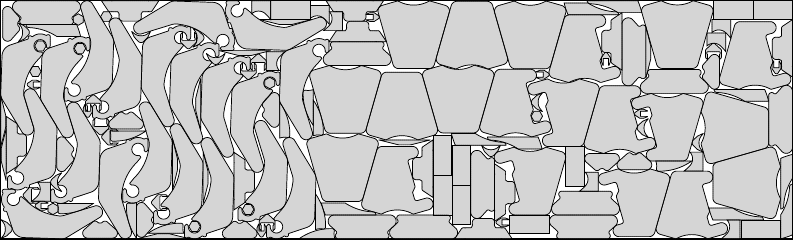}
        \caption{\texttt{GARDEYN7}}
    \end{subfigure}
    \hfill
    \begin{subfigure}[t]{0.33\linewidth}
        \centering
        \includegraphics[height=\subfigheight]{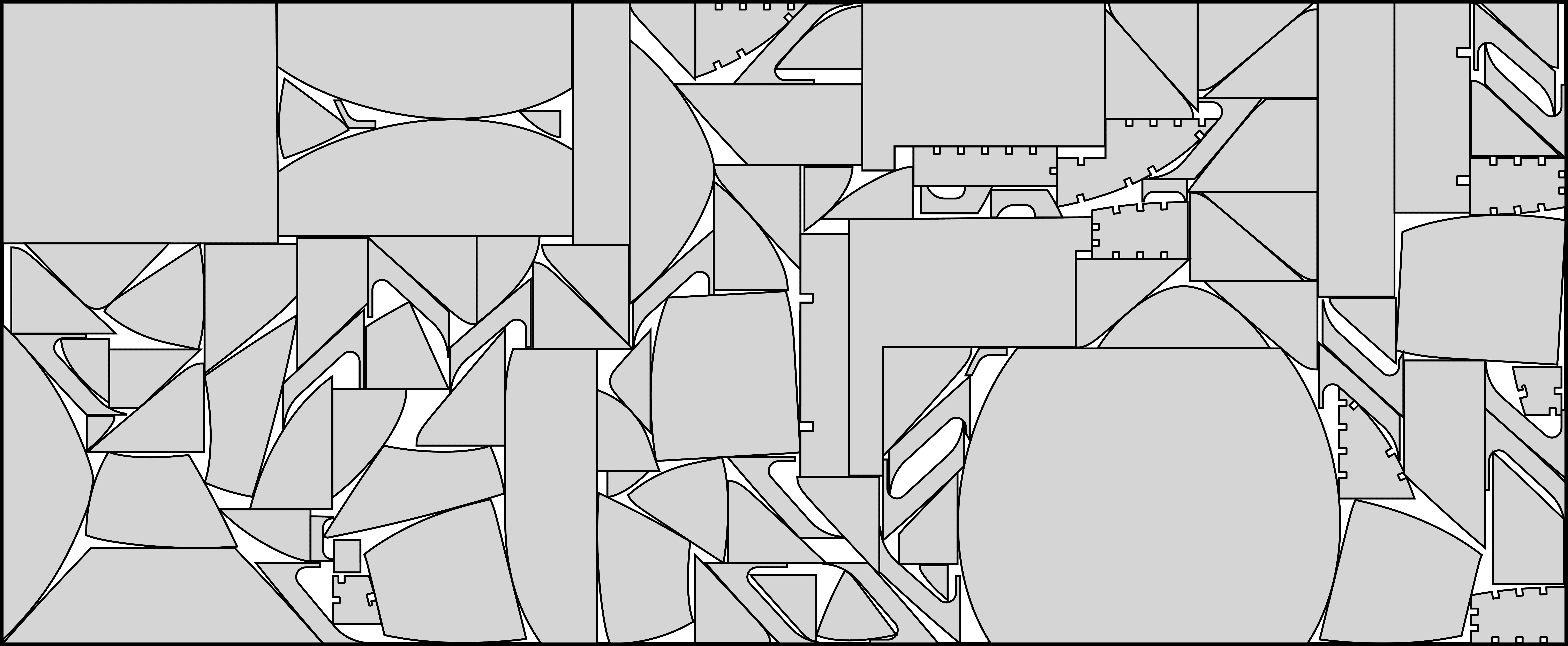}
        \caption{\texttt{GARDEYN8}}
    \end{subfigure}
    \hfill
    \begin{subfigure}[t]{0.2\linewidth}
        \hfill
        \includegraphics[height=\subfigheight]{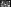}
        \caption{\texttt{GARDEYN9}}
    \end{subfigure}

    \caption{Ten new benchmark instances for the 2DISPP. The visualizations are solutions produced by \texttt{sparrow}.}
    \label{fig:new_instances}
\end{figure}

Figure \ref{fig:new_instances} presents this new suite instances for the 2DISPP.
Instances \texttt{GARDEYN0} - \texttt{3} have been derived from \citet{corentin_lallier_2022_7030786}, a large dataset of nesting jobs originating from fashion industry clients of Lectra\footnote{\url{https://lectra.com/en}}.
Instances \texttt{GARDEYN4} - \texttt{8} are based on shapes from metalworking and shipbuilding applications, provided by AlmaCAM\footnote{\url{https://almacam.com/}}.
Finally, \texttt{GARDEYN9} represents a direct-to-film printing context, contributed by Tetrinest\footnote{\url{https://tetrinest.com/}}.

These new instances revealed a serious blind spot in the traditional academic benchmarks.
Highly homogeneous instances -- such as \texttt{GARDEYN4}, \texttt{5}, and \texttt{7} -- often contain subsets of items that can be clustered together in a compact way, and repeated many times throughout the strip.
While \texttt{sparrow} can generate these compact patterns at a local scale, it lacks a mechanism to repeat them and is therefore unable to exploit the inherent `structure' or `regularity' made possible by the homogeneity of the items.
This characteristic is prevalent in real-world applications, but was not captured by the traditional academic benchmarks.

Extending \texttt{sparrow} with the ability to exploit this inherent structure/regularity is a promising direction for future advances, as we are confident vastly superior solutions are attainable, especially for the aforementioned homogeneous instances.

\section{Conclusions and reflections} \label{section:conclusion}
This paper introduced a heuristic for the 2D irregular strip packing problem (2DISPP) that redefines both the academic and open-source state of the art.
Our implementation, \texttt{sparrow}, is publicly available at: \url{https://github.com/JeroenGar/sparrow}.
The core idea of the algorithm is to decompose the 2DISPP into a sequence of feasibility problems, where the strip is converted into a container with fixed dimensions and the items are temporarily allowed to collide with each other.
A local-search algorithm iteratively moves colliding items around in an attempt to gradually resolve them all and eventually reach feasibility.
Although we focus on the 2DISPP, most of the heuristic's components operate on the feasibility problem and can thus easily be incorporated into solution strategies for other 2D nesting problems such as bin packing or knapsack problems.

The main contributions of this work are fourfold: 
(i) a reproducible and substantial improvement in solution quality across both academic and open-source contexts,
(ii) a highly-optimized and robust implementation ready for deployment in practice or as a base for future research,
(iii) a number of foundational algorithmic ideas that can be iteratively improved upon or adapted to other problem variants and
(iv) a new set of academic benchmark instances designed to properly reflect real-world nesting applications and uncover weaknesses of the approach.

To be frank, the scale of improvements with respect to the previous state of the art demonstrated in the computational experiments (Section \ref{section:experiments}) should not have been possible on such a well-studied problem using benchmark instances that are over 25 years old.
For the most challenging instance, \texttt{SWIM}, \texttt{sparrow}'s expected and best solution are both $\sim$5\% more compact than the previous best results.
Such a margin of improvement confirms that stagnation with respect to 2DISPP research was not a result of it already being solved to near optimality.

Progress in this field has instead been held back for far too long by the issues we outlined in the introduction: the persistence of high barrier to entry and a severe lack of reproducibility.
In an alternate scenario -- where this paper were published without its open-source implementation -- the method we propose would have simply continued that trend and been nearly impossible to reproduce in practice.
Even worse, by raising the performance bar while simultaneously leaving others unable to meaningfully verify or build upon our work, we could have inadvertently initiated a new decade of stagnation.

Reproducibility is a cornerstone of the scientific method.
Without it, researchers are left in an unhealthy and inefficient environment: forced to reinvent the wheel without the means to increment on prior work or even validate its results.
We strongly believe that an academic paper coupled with a high-quality and open-source implementation delivers value significantly exceeding the sum of its parts.
Thankfully, the ethos we are advocating is not swimming against the current. 
The field of Operations Research is undergoing a gradual cultural shift toward open-source practices \citep{petropoulos2024operational}. 
\citet{vidal2022hybrid}, for example, recently published an open-source implementation of their state-of-the-art algorithm for the capacitated vehicle routing problem.
Many of the concerns we raise here are in fact echoing sentiments already expressed by \citet{vidal2022hybrid}.

We hope this paper and \texttt{sparrow} will serve as catalysts for continued advances in 2D nesting research as well as deterrents to non-reproducible research practices that have historically hindered progress.

\section*{Acknowledgements}
The authors wish to explicitly thank Luke Connolly (Connolly Editorial) for editorial consultation and Martial Luyts (LStat, KU Leuven) for statistical consultation.
They also thank Teo Mazars, Luc Libralesso, Thomas Piotaix, Corbin Linder, and Léo Gilbert for their contributions to the creation of the new benchmark instances.
This research was supported by the Research Foundation — Flanders (FWO) under grant numbers 1S71222N and K804824N.

\bibliography{bibliography.bib}

\end{document}